\documentclass[twocolumn, aps, superscriptaddress, groupedaddress]{revtex4}

\usepackage{mathtools}
 
\usepackage{amsmath,amsthm,amsfonts,amssymb}  
\usepackage{fontenc}
\usepackage[all]{xy}
\usepackage{graphicx,subfigure,multirow}
\usepackage{tikz}
\usetikzlibrary{shapes, arrows, shadows} 
\usepackage[latin1]{inputenc}
\usepackage[scaled]{helvet}
\usepackage[toc, page]{appendix}
\usepackage{graphicx}				
\usepackage{amssymb}
\newtheorem{definition}{Definition}
\newtheorem{theorem}{Theorem}
\newtheorem{lemma}{Lemma}

\usepackage{tikz}
\pgfdeclarelayer{nodelayer}
\pgfdeclarelayer{edgelayer}
\pgfsetlayers{edgelayer,nodelayer,main}
\tikzstyle{none}=[inner sep=0pt]
\tikzstyle{new}=[inner sep=2pt]
\usetikzlibrary{shapes, arrows, shadows}
\usetikzlibrary{circuits.ee.IEC}
\usetikzlibrary{decorations.pathmorphing}
\usetikzlibrary{shapes.geometric}

\tikzstyle{env}=[copoint,regular polygon rotate=0,minimum width=0.2cm, fill=black]

\tikzstyle{probs}=[shape=semicircle,fill=white,draw=black,shape border rotate=180,minimum width=1.2cm]

\usetikzlibrary{circuits.ee.IEC}
\usetikzlibrary{decorations.pathmorphing}
\usetikzlibrary{shapes.geometric}

\tikzstyle{wavy}=[decorate,decoration={snake, segment length=1mm, amplitude=0.3mm}]
\tikzstyle{mopoint}=[shape=semicircle, fill=white,draw=black,shape border rotate=180,scale =0.75]
\tikzstyle{mocopoint}=[shape=semicircle, fill=white,draw=black,minimum width = 0.9cm, scale =0.75, xscale=0.7]
\tikzstyle{cpoint}=[shape=semicircle, fill=white,draw=black,minimum width = 0.9cm, scale =0.75, xscale=1, yscale=0.7, shape border rotate = 90,font=\fontsize{14}{16}\selectfont]
\tikzstyle{cocpoint}=[shape=semicircle, fill=white,draw=black,minimum width = 0.9cm, scale =0.75, xscale=1, yscale=0.7, shape border rotate = 270,font=\fontsize{14}{16}\selectfont]
\tikzstyle{every picture}=[baseline=-0.25em,scale=0.5]
\tikzstyle{dotpic}=[] % for backwards-compatibility
\tikzstyle{diredges}=[every to/.style={diredge}]
\tikzstyle{math matrix}=[matrix of math nodes,left delimiter=(,right delimiter=),inner sep=2pt,column sep=1em,row sep=0.5em,nodes={inner sep=0pt},text height=1.5ex, text depth=0.25ex]

\tikzstyle{inline text}=[text height=1.5ex, text depth=0.25ex,yshift=0.5mm]
\tikzstyle{label}=[font=\footnotesize,text height=1.5ex, text depth=0.25ex,yshift=0.5mm]
\tikzstyle{left label}=[label,anchor=east,xshift=1.5mm]
\tikzstyle{right label}=[label,anchor=west,xshift=-1.5mm]

\tikzstyle{braceedge}=[decorate,decoration={brace,amplitude=2mm,raise=-1mm}]
\tikzstyle{small braceedge}=[decorate,decoration={brace,amplitude=1mm,raise=-1mm}]

\tikzstyle{doubled}=[line width=2pt] % set the line width for all doubled (quantum) maps/wires
\tikzstyle{boldedge}=[doubled,shorten <=-0.17mm,shorten >=-0.17mm]

\tikzstyle{boldedgedashed}=[very thick,dashed,shorten <=-0.17mm,shorten >=-0.17mm]
\tikzstyle{vboldedgedashed}=[doubled,dashed,shorten <=-0.17mm,shorten >=-0.17mm]
\tikzstyle{left hook arrow}=[left hook-latex]
\tikzstyle{right hook arrow}=[right hook-latex]
\tikzstyle{sembracket}=[line width=0.5pt,shorten <=-0.07mm,shorten >=-0.07mm]

\tikzstyle{causal edge}=[->,thick,gray]
\tikzstyle{causal nondir}=[thick,gray]
\tikzstyle{timeline}=[thick,gray, dashed]

\tikzstyle{cedge}=[<->,thick,gray!70!white]

\tikzstyle{empty diagram}=[draw=gray!40!white,dashed,shape=rectangle,minimum width=1cm,minimum height=1cm]
\tikzstyle{empty diagram small}=[draw=gray!50!white,dashed,shape=rectangle,minimum width=0.6cm,minimum height=0.5cm]

\tikzstyle{dot}=[inner sep=0.7mm,minimum width=0pt,minimum height=0pt,draw,shape=circle]
\tikzstyle{ddot}=[inner sep=0.7mm,doubled, minimum width=2.5mm,minimum height=2.5mm,draw,shape=circle]

\tikzstyle{black dot}=[dot,fill=black]
\tikzstyle{white dot}=[dot,fill=white]
\tikzstyle{green dot}=[white dot] % for backwards-compatibility
\tikzstyle{gray dot}=[dot,fill=gray!40!white]
\tikzstyle{red dot}=[gray dot] % for backwards-compatibility

\tikzstyle{black ddot}=[ddot,fill=black]
\tikzstyle{white ddot}=[ddot,fill=white]
\tikzstyle{gray ddot}=[ddot,fill=gray!40!white]

\tikzstyle{gray edge}=[gray!40!white]

\tikzstyle{small dot}=[inner sep=0.4mm,minimum width=0pt,minimum height=0pt,draw,shape=circle]

\tikzstyle{small black dot}=[small dot,fill=black]
\tikzstyle{small white dot}=[small dot,fill=white]
\tikzstyle{small gray dot}=[small dot,fill=gray!40!white]

\tikzstyle{causal dot}=[inner sep=0.4mm,minimum width=0pt,minimum height=0pt,draw=white,shape=circle,fill=gray!40!white]

\tikzstyle{white phase dot}=[dot,fill=white]
\tikzstyle{white phase ddot}=[ddot,fill=white]

\tikzstyle{gray phase dot}=[dot,fill=gray!40!white]
\tikzstyle{gray phase ddot}=[ddot,fill=gray!40!white]
\tikzstyle{grey phase dot}=[gray phase dot]
\tikzstyle{grey phase ddot}=[gray phase ddot]

\tikzstyle{cnot}=[fill=white,shape=circle,inner sep=-1.4pt]
\tikzstyle{hadamard}=[square box,inner sep=0 pt,font=\tiny\sf,minimum height=3mm,minimum width=3mm]
\tikzstyle{dhadamard}=[hadamard,doubled]
\tikzstyle{antipode}=[white dot,inner sep=0.3mm,font=\footnotesize]

\tikzstyle{scalar}=[diamond,draw,inner sep=0.5pt,font=\small]
\tikzstyle{dscalar}=[diamond,doubled, draw,inner sep=0.5pt,font=\small]

\tikzstyle{small box}=[rectangle,inline text,fill=white,draw,minimum height=5mm,yshift=-0.5mm,minimum width=5mm,font=\small]
\tikzstyle{small gray box}=[small box,fill=gray!30]
\tikzstyle{medium box}=[rectangle,inline text,fill=white,draw,minimum height=5mm,yshift=-0.5mm,minimum width=10mm,font=\small]
\tikzstyle{square box}=[small box] % for backwards-compatibility
\tikzstyle{medium gray box}=[small box,fill=gray!30]
\tikzstyle{large box}=[rectangle,inline text,fill=white,draw,minimum height=5mm,yshift=-0.5mm,minimum width=15mm,font=\small]
\tikzstyle{large gray box}=[small box,fill=gray!30]

\tikzstyle{point}=[regular polygon,regular polygon sides=3,draw,scale=0.75,inner sep=-0.5pt,minimum width=9mm,fill=white,regular polygon rotate=180]
\tikzstyle{copoint}=[regular polygon,regular polygon sides=3,draw,scale=0.75,inner sep=-0.5pt,minimum width=9mm,fill=white]
\tikzstyle{dpoint}=[point,doubled]
\tikzstyle{dcopoint}=[copoint,doubled]

\tikzstyle{tinypoint}=[regular polygon,regular polygon sides=3,draw,scale=0.55,inner sep=-0.15pt,minimum width=6mm,fill=white,regular polygon rotate=180]

\tikzstyle{white point}=[point]
\tikzstyle{green point}=[white point] % for backwards-compatibility
\tikzstyle{white copoint}=[copoint]
\tikzstyle{gray point}=[point,fill=gray!40!white]
\tikzstyle{gray dpoint}=[gray point,doubled]
\tikzstyle{red point}=[gray point] % for backwards-compatibility
\tikzstyle{gray copoint}=[copoint,fill=gray!40!white]
\tikzstyle{gray dcopoint}=[gray copoint,doubled]

\tikzstyle{tiny gray point}=[tinypoint,fill=gray!40!white]

\tikzstyle{diredge}=[->]
\tikzstyle{rdiredge}=[<-]
\tikzstyle{thickdiredge}=[->, very thick]
\tikzstyle{pointer edge}=[->,very thick,gray]
\tikzstyle{pointer edge part}=[very thick,gray]
\tikzstyle{dashed edge}=[dashed]
\tikzstyle{thick dashed edge}=[very thick,dashed]
\tikzstyle{thick gray dashed edge}=[thick dashed edge,gray!90]
\tikzstyle{thick map edge}=[very thick,|->]

\makeatletter
\newcommand{\boxshape}[3]{%
\pgfdeclareshape{#1}{
\inheritsavedanchors[from=rectangle] % this is nearly a rectangle
\inheritanchorborder[from=rectangle]
\inheritanchor[from=rectangle]{center}
\inheritanchor[from=rectangle]{north}
\inheritanchor[from=rectangle]{south}
\inheritanchor[from=rectangle]{west}
\inheritanchor[from=rectangle]{east}
\backgroundpath{% this is new
\southwest \pgf@xa=\pgf@x \pgf@ya=\pgf@y
\northeast \pgf@xb=\pgf@x \pgf@yb=\pgf@y

\@tempdima=#2
\@tempdimb=#3

\pgfpathmoveto{\pgfpoint{\pgf@xa - 5pt + \@tempdima}{\pgf@ya}}
\pgfpathlineto{\pgfpoint{\pgf@xa - 5pt - \@tempdima}{\pgf@yb}}
\pgfpathlineto{\pgfpoint{\pgf@xb + 5pt + \@tempdimb}{\pgf@yb}}
\pgfpathlineto{\pgfpoint{\pgf@xb + 5pt - \@tempdimb}{\pgf@ya}}
\pgfpathlineto{\pgfpoint{\pgf@xa - 5pt + \@tempdima}{\pgf@ya}}
\pgfpathclose
}
}}

\boxshape{NEbox}{0pt}{5pt}
\boxshape{SEbox}{0pt}{-5pt}
\boxshape{NWbox}{5pt}{0pt}
\boxshape{SWbox}{-5pt}{0pt}
\boxshape{EBox}{-3pt}{3pt}
\boxshape{WBox}{3pt}{-3pt}
\makeatother

\tikzstyle{cloud}=[shape=cloud,draw,minimum width=1.5cm,minimum height=1.5cm]

\tikzstyle{map}=[draw,shape=NEbox,inner sep=2pt,minimum height=6mm,fill=white]
\tikzstyle{mapdag}=[draw,shape=SEbox,inner sep=2pt,minimum height=6mm,fill=white]
\tikzstyle{mapadj}=[draw,shape=SEbox,inner sep=2pt,minimum height=6mm,fill=white]
\tikzstyle{maptrans}=[draw,shape=SWbox,inner sep=2pt,minimum height=6mm,fill=white]
\tikzstyle{mapconj}=[draw,shape=NWbox,inner sep=2pt,minimum height=6mm,fill=white]

\tikzstyle{dbox}=[draw,doubled,shape=rectangle,inner sep=2pt,minimum height=6mm,minimum width=6mm,fill=white]
\tikzstyle{dmap}=[draw,doubled,shape=NEbox,inner sep=2pt,minimum height=6mm,fill=white]
\tikzstyle{dmapdag}=[draw,doubled,shape=SEbox,inner sep=2pt,minimum height=6mm,fill=white]
\tikzstyle{dmapadj}=[draw,doubled,shape=SEbox,inner sep=2pt,minimum height=6mm,fill=white]
\tikzstyle{dmaptrans}=[draw,doubled,shape=SWbox,inner sep=2pt,minimum height=6mm,fill=white]
\tikzstyle{dmapconj}=[draw,doubled,shape=NWbox,inner sep=2pt,minimum height=6mm,fill=white]

\tikzstyle{ddmap}=[draw,doubled,dashed,shape=NEbox,inner sep=2pt,minimum height=6mm,fill=white]
\tikzstyle{ddmapdag}=[draw,doubled,dashed,shape=SEbox,inner sep=2pt,minimum height=6mm,fill=white]
\tikzstyle{ddmapadj}=[draw,doubled,dashed,shape=SEbox,inner sep=2pt,minimum height=6mm,fill=white]
\tikzstyle{ddmaptrans}=[draw,doubled,dashed,shape=SWbox,inner sep=2pt,minimum height=6mm,fill=white]
\tikzstyle{ddmapconj}=[draw,doubled,dashed,shape=NWbox,inner sep=2pt,minimum height=6mm,fill=white]

\boxshape{sNEbox}{0pt}{3pt}
\boxshape{sSEbox}{0pt}{-3pt}
\boxshape{sNWbox}{3pt}{0pt}
\boxshape{sSWbox}{-3pt}{0pt}
\tikzstyle{smap}=[draw,shape=sNEbox,fill=white]
\tikzstyle{smapdag}=[draw,shape=sSEbox,fill=white]
\tikzstyle{smapadj}=[draw,shape=sSEbox,fill=white]
\tikzstyle{smaptrans}=[draw,shape=sSWbox,fill=white]
\tikzstyle{smapconj}=[draw,shape=sNWbox,fill=white]

\tikzstyle{dsmap}=[draw,dashed,shape=sNEbox,fill=white]
\tikzstyle{dsmapdag}=[draw,dashed,shape=sSEbox,fill=white]
\tikzstyle{dsmaptrans}=[draw,dashed,shape=sSWbox,fill=white]
\tikzstyle{dsmapconj}=[draw,dashed,shape=sNWbox,fill=white]

\boxshape{mNEbox}{0pt}{10pt}
\boxshape{mSEbox}{0pt}{-10pt}
\boxshape{mNWbox}{10pt}{0pt}
\boxshape{mSWbox}{-10pt}{0pt}
\tikzstyle{mmap}=[draw,shape=mNEbox]
\tikzstyle{mmapdag}=[draw,shape=mSEbox]
\tikzstyle{mmaptrans}=[draw,shape=mSWbox]
\tikzstyle{mmapconj}=[draw,shape=mNWbox]

\tikzstyle{mmapgray}=[draw,fill=gray!40!white,shape=mNEbox]
\tikzstyle{smapgray}=[draw,fill=gray!40!white,shape=sNEbox]

\makeatletter
\pgfdeclareshape{cornerpoint}{
\inheritsavedanchors[from=rectangle] % this is nearly a rectangle
\inheritanchorborder[from=rectangle]
\inheritanchor[from=rectangle]{center}
\inheritanchor[from=rectangle]{north}
\inheritanchor[from=rectangle]{south}
\inheritanchor[from=rectangle]{west}
\inheritanchor[from=rectangle]{east}
\backgroundpath{% this is new
\southwest \pgf@xa=\pgf@x \pgf@ya=\pgf@y
\northeast \pgf@xb=\pgf@x \pgf@yb=\pgf@y

\pgfmathsetmacro{\pgf@shorten@left}{\pgfkeysvalueof{/tikz/shorten left}}
\pgfmathsetmacro{\pgf@shorten@right}{\pgfkeysvalueof{/tikz/shorten right}}

\pgfpathmoveto{\pgfpoint{0.5 * (\pgf@xa + \pgf@xb)}{\pgf@ya - 5pt}}
\pgfpathlineto{\pgfpoint{\pgf@xa - 8pt + \pgf@shorten@left}{\pgf@yb - 1.5 * \pgf@shorten@left}}
\pgfpathlineto{\pgfpoint{\pgf@xa - 8pt + \pgf@shorten@left}{\pgf@yb}}
\pgfpathlineto{\pgfpoint{\pgf@xb + 8pt - \pgf@shorten@right}{\pgf@yb}}
\pgfpathlineto{\pgfpoint{\pgf@xb + 8pt - \pgf@shorten@right}{\pgf@yb - 1.5 * \pgf@shorten@right}}
\pgfpathclose
}
}

\pgfdeclareshape{cornercopoint}{
\inheritsavedanchors[from=rectangle] % this is nearly a rectangle
\inheritanchorborder[from=rectangle]
\inheritanchor[from=rectangle]{center}
\inheritanchor[from=rectangle]{north}
\inheritanchor[from=rectangle]{south}
\inheritanchor[from=rectangle]{west}
\inheritanchor[from=rectangle]{east}
\backgroundpath{% this is new
\southwest \pgf@xa=\pgf@x \pgf@ya=\pgf@y
\northeast \pgf@xb=\pgf@x \pgf@yb=\pgf@y

\pgfmathsetmacro{\pgf@shorten@left}{\pgfkeysvalueof{/tikz/shorten left}}
\pgfmathsetmacro{\pgf@shorten@right}{\pgfkeysvalueof{/tikz/shorten right}}

\pgfpathmoveto{\pgfpoint{0.5 * (\pgf@xa + \pgf@xb)}{\pgf@yb + 5pt}}
\pgfpathlineto{\pgfpoint{\pgf@xa - 8pt + \pgf@shorten@left}{\pgf@ya + 1.5 * \pgf@shorten@left}}
\pgfpathlineto{\pgfpoint{\pgf@xa - 8pt + \pgf@shorten@left}{\pgf@ya}}
\pgfpathlineto{\pgfpoint{\pgf@xb + 8pt - \pgf@shorten@right}{\pgf@ya}}
\pgfpathlineto{\pgfpoint{\pgf@xb + 8pt - \pgf@shorten@right}{\pgf@ya + 1.5 * \pgf@shorten@right}}
\pgfpathclose
}
}

\makeatother

\pgfkeyssetvalue{/tikz/shorten left}{0pt}
\pgfkeyssetvalue{/tikz/shorten right}{0pt}

\tikzstyle{kpoint common}=[draw,fill=white,inner sep=1pt,minimum height=4mm]
\tikzstyle{kpoint}=[shape=cornerpoint,shorten left=5pt,kpoint common]
\tikzstyle{kpoint adjoint}=[shape=cornercopoint,shorten left=5pt,kpoint common]
\tikzstyle{kpoint conjugate}=[shape=cornerpoint,shorten right=5pt,kpoint common]
\tikzstyle{kpoint transpose}=[shape=cornercopoint,shorten right=5pt,kpoint common]
\tikzstyle{kpoint symm}=[shape=cornerpoint,shorten left=5pt,shorten right=5pt,kpoint common]

\tikzstyle{kpointdag}=[kpoint adjoint]
\tikzstyle{kpointadj}=[kpoint adjoint]
\tikzstyle{kpointconj}=[kpoint conjugate]
\tikzstyle{kpointtrans}=[kpoint transpose]

\tikzstyle{dkpoint}=[kpoint,doubled]
\tikzstyle{dkpointdag}=[kpoint adjoint,doubled]
\tikzstyle{dkcopoint}=[kpoint adjoint,doubled]
\tikzstyle{dkpointadj}=[kpoint adjoint,doubled]
\tikzstyle{dkpointconj}=[kpoint conjugate,doubled]
\tikzstyle{dkpointtrans}=[kpoint transpose,doubled]

\tikzstyle{kscalar}=[kpoint common, shape=EBox, inner xsep=-1pt, inner ysep=3pt,font=\small]
\tikzstyle{kscalarconj}=[kpoint common, shape=WBox, inner xsep=-1pt, inner ysep=3pt,font=\small]

 \tikzstyle{upground}=[circuit ee IEC,thick,ground,rotate=90,scale=2.5]
 \tikzstyle{downground}=[circuit ee IEC,thick,ground,rotate=-90,scale=2.5]
 \tikzstyle{bigground}=[regular polygon,regular polygon sides=3,draw=gray,scale=0.50,inner sep=-0.5pt,minimum width=10mm,fill=gray]
\tikzstyle{arrs}=[-latex,font=\small,auto]
\tikzstyle{arrow plain}=[arrs]
\tikzstyle{arrow dashed}=[dashed,arrs]
\tikzstyle{arrow bold}=[very thick,arrs]
\tikzstyle{arrow hide}=[draw=white!0,-]
\tikzstyle{arrow reverse}=[latex-]
\tikzstyle{cdnode}=[]

\tikzstyle{slit}=[line width=2]
\tikzstyle{block}=[line width=4,gray,line cap=round]
\tikzstyle{screen}=[line width=4,black,line cap=round]
\tikzstyle{di}=[diamond,draw,inner sep=0.5pt,font=\small, minimum size = .5cm]
\tikzstyle{sbox}=[rectangle,draw]
\tikzstyle{mirror}=[line width=2,black]
\tikzstyle{trace}=[circuit ee IEC,thick,ground,rotate=0,scale=2]
\tikzstyle{traceState}=[circuit ee IEC,thick,ground,rotate=180,scale=2]
\tikzstyle{detEff}=[circuit ee IEC,thick,ground,rotate=180,scale=1.4]
\tikzstyle{maxMix}=[circuit ee IEC,thick,ground,scale=1.4]
\tikzstyle{particlePath}=[line width=2,gray!40, line cap =round]
\usepackage{fp}
\usetikzlibrary{calc,decorations.pathmorphing,decorations.text,fixedpointarithmetic}

\begin{document}

\title{The computational landscape of general physical theories}   
\author{Jonathan Barrett}
\affiliation{University of Oxford, Department of Computer Science, Wolfson Building, Parks Road, Oxford OX1 3QD, UK}
\author{Niel de Beaudrap} 
\affiliation{University of Oxford, Department of Computer Science, Wolfson Building, Parks Road, Oxford OX1 3QD, UK}
\author{Matty~J. Hoban}
\affiliation{University of Oxford, Department of Computer Science, Wolfson Building, Parks Road, Oxford OX1 3QD, UK}
\affiliation{Department of Computing, Goldsmiths, University of London, New Cross, London SE14 6NW, UK}
\author{Ciar{\'a}n~M. Lee}
\email{ciaran.lee@ucl.ac.uk}
\affiliation{Department of Physics and Astronomy, University College London, Gower Street, London WC1E 6BT, UK}

\begin{abstract} 
%The emergence of quantum computers has challenged long-held beliefs about what is efficiently computable given our current physical theories. However, going back to the work of Abrams and Lloyd, changing one aspect of quantum theory can result in yet more dramatic increases in computational power, as well as violations of fundamental physical principles. This raises the question of whether there exists a well-defined physical theory which exceeds the power of quantum computation, but still satisfies fundamental physical principles. In the current work we focus on efficient computation within a framework of general physical theories that make good operational sense. We explicitly construct a post-quantum theory that satisfies the principles of causality (roughly, no signalling from the future) and tomographic locality (roughly, multipartite states can be characterised by local measurements) which can efficiently simulate any computation in this framework, including quantum computation. In analogy with the study of non-local quantum correlations, this leads us to question what physical principles recover the power of quantum computing. 

There is good evidence that quantum computers are more powerful than classical computers, and that various simple modifications of quantum theory yield computational power that is dramatically greater still. However, these modifications also violate fundamental physical principles. This raises the question of whether there exists a physical theory, allowing computation more powerful than quantum, but which still respects those fundamental physical principles. Prior work by two of us introduced this question within a suitable framework for theories that make good operational sense, and showed that in any theory satisfying tomographic locality, the class of problems that can be solved efficiently is contained in the complexity class $\bold{AWPP}$. Here, we show that this bound is tight, in the sense that there exists a theory, satisfying tomographic locality, as well as a basic principle of causality, which can efficiently decide everything in $\bold{AWPP}$. Hence this theory can efficiently simulate any computation in this framework, including quantum computation.

%Thus the class $\bold{AWPP}$ has a natural physical interpretation: it is precisely the class of problems that can be solved efficiently in tomographically-local theories. The result depends on a lemma, which characterizes $\bold{AWPP}$ precisely as the class of problems solvable by quasi-stochastic Turing machines.
\end{abstract} 

\maketitle 

There is ever-growing evidence that quantum computers are more powerful than classical computers \cite{arkhipov,shor,IQC,IQC1}. However, an understanding of the source of this power remains elusive. Many features of quantum mechanics have been posited as the origin of this so-called ``speed-up" \cite{Magic,Vidal,Hoban-2011,discord,Interference-speed-up} but the debate is far from resolved \cite{Vedral, anti-discord, entanglement}. In recent years, one way of examining this power has been to ask how the computational power changes as features of quantum theory are altered. Beginning with the work of Abrams and Lloyd, it was shown that allowing more exotic transformations in quantum theory can result in easily solving hard problems \cite{abrams}. This motivates the speculation that quantum theory is an ``island" within the space of all possible theories; alter quantum mechanics and we obtain dramatic consequences \cite{aaronsonpost}. 
 
Another possibility is that our understanding of computation in possible physical theories is couched too much in the language of quantum theory. For example, it could be entirely possible to have a theory that has the same computational power as quantum theory but barely resembles it. We thus require an abstract framework in which to study the power of computation, where quantum and classical computation are special cases.  

The study of operational theories provides us with a suitable framework for the study of information processing based on operational principles \cite{Pavia1,Hardy-2011,Barrett-2007,Masanes,LS-2017}. That is, we can make statements about the limits and power of information processing without referring explicitly to quantum theory. Some features thought unique to quantum theory (as opposed to classical physics) can be seen to be ubiquitous within these theories. For example, given some fundamental properties that reasonable operational theories should satisfy, a no-broadcasting theorem holds in any non-classical theory \cite{Broadcast}.  This then begs the question of what fundamental principles uniquely single out quantum physics from these myriad possibilities. Indeed, starting from various frameworks of operational theories there have been many derivations of quantum theory from information theoretic principles (e.g.~Refs.~\cite{Pavia2,Hardy-2011,Masanes2}).

In Refs.~\cite{LB-2014, LH-2015, LH-2016}, a circuit-based model of computation is defined and studied in the context of a broad operationally-defined framework for physical theories. Informally, a theory in this framework specifies a set of laboratory devices that can be connected together to form experiments, and assigns probabilities to experimental outcomes. Whilst many such theories may not correspond to descriptions of our physical world, they nevertheless make good operational sense, and allow one to systematically assess how computational power depends on the underlying physical theory. 
 
One can identify physical principles that theories may or may not satisfy, such as causality (no signalling from future to past), or tomographic locality (local measurements suffice for tomography of joint states). Ref.~\cite{LB-2014} shows that for theories satisfying tomographic locality, whether or not causality is satisfied, computational problems that can be solved efficiently are contained in the classical complexity class $\bold{AWPP}$---a bound first proved for the quantum case by Fortnow and Rogers \cite{AWPP}. 

Ref.~\cite{LB-2014} leaves open the question of whether the bound is tight, in the sense that there exists a theory that could solve all problems in $\bold{AWPP}$. Such a theory would have computational power beyond that which we expect from quantum mechanics and could simulate any quantum computation. In this paper we resolve this open problem and show that there does indeed exist a non-quantum theory, satisfying both tomographic locality and causality, which can decide everything in $\bold{AWPP}$. We may consider this theory as a ``foil" theory, used to deepen our understanding of the limitations of quantum computers. This foil theory is constructed from a computational model using quasi-probabilities, i.e.~an affine combination of weights assigned to particular events. This motivates the study of what minimal set of information principles recover the power of quantum computation.

%Section~\ref{framework} presents a circuit model of computation in the setting of general physical theories, defines efficient computation in such a theory, and gives computational complexity bounds for this notion of computation. Section~\ref{upperbound} describes the theory that attains the bound of $\bold{AWPP}$, generating the theory from a circuit construction related to a family of computational models called Affine Turing Machines. Section~\ref{upperbound}, gives some computational complexity evidence that quantum computers will not achieve $\bold{AWPP}$. Sections~\ref{main proof} and \ref{proof of promise} contain the details of our constructions and results. Finally, Section~\ref{discussion}, concludes with some remarks about the problem of determining the computational complexity of quantum computers from basic informational principles, in analogy with the characterization of quantum non-locality \cite{Popescu-review,PR-van-Dam,Henson-2015,Almost-quantum,local-orthogonality,information-causality}.

\section*{Results}\label{framework}

\subsection{Operational theories}

The fundamental goal of any physical theory is to provide a consistent account of experimental data. This constitutes the core idea underlying the framework of operational theories \cite{hardy2001quantum, Barrett-2007, Pavia1, Pavia2, Hardy-2011, de2012deriving, masanes2011derivation, LB-2014}, where the primitive notions are operational in nature. 

A theory in this framework specifies a set of \emph{laboratory devices}, which can be connected together in certain ways, and assigns probabilities to different experimental outcomes. A laboratory device comes equipped with \emph{input ports}, \emph{output ports}, and a \emph{classical pointer}, where roughly speaking, one may think of physical systems passing into input ports and emerging from output ports, with the pointer indicating an experimental outcome. Each input and output port has an associated \emph{type}. We will often denote types $A,B,C\ldots$, and use $X$ or $Y$ to stand for generic types.  Experiments correspond to circuits, which are formed by connecting output ports of devices to input ports of other devices in such a way that types match. By assumption, the circuit corresponding to a valid experiment must be acyclic, and closed, meaning that there are no unconnected input or output ports. When an experiment is run, each pointer comes to rest in a final position, with these pointer positions constituting jointly the outcome of the experiment. For any circuit corresponding to an allowed experiment, the theory must define a joint probability distribution over pointer positions for all devices in the circuit.

Laboratory devices include \emph{preparation devices}, which have no input ports, and \emph{measurement devices}, which have no output ports. Each use of a preparation device outputs a physical system in some particular \emph{state}, where the state is determined by the variety of device used and the position attained by the pointer on the device on that run. A measurement device can be thought of as implementing a destructive measurement, since no system emerges, with the outcome denoted by the pointer position. Each outcome corresponds to an \emph{effect}. Given a device with both input and output ports, a system may pass through in such a way that its state is altered. The change in the state is non-deterministic in the sense that the change applied is indicated by the position of the pointer. When a device has both input and output ports, each pointer position corresponds to a \emph{transformation}.

For the formal development of operational theories, see for example Refs.~\cite{hardy2001quantum, Barrett-2007, Pavia1, Pavia2, Hardy-2011, de2012deriving, masanes2011derivation, LB-2014}. Here, rather than present an axiomatic derivation, we simply summarize the resulting mathematical structure. 

Each system type $X$ can be associated with a real vector space $V_X$, such that a state of the system is a vector in $V_X$. In this work, it is assumed throughout that for each type of system, $V_X$ is finite dimensional. Types are closed under parallel composition, hence given a system of type $X$ and a system of type $Y$, there is a composite system, whose type can be denoted $XY$. The theories that we are interested in satisfy the principle of  \emph{tomographic locality} \cite{Barrett-2007, Pavia1}, which says that multipartite states can be uniquely specified by the joint probabilities for the outcomes of measurements performed locally on each component system. This implies that the vector space associated with a composite system is the tensor product of the vector spaces associated with the component systems: i.e., the vector space associated with the composite $XY$ is $V_{XY}=V_X\otimes V_Y$. A state of the composite is a direct product state if it is of the form $|s)_{XY} = |s)_X \otimes |s)_Y$.

A transformation, with input type $X$ and output type $Y$, is a linear map from $V_X$ to $V_Y$. Given a composite system of type $XY$, the parallel action of transformation $T_X$ on the type $X$ subsystem, and transformation $T_Y$ on the type $Y$ subsystem, is given by a transformation $T_{XY} = T_X\otimes T_Y$. An effect on a system of type $X$ is a linear map from $V_X$ to the real numbers, i.e., an effect is an element of the dual space. Consider a composite system of type $XY$, and suppose that local measurements are performed. If a particular outcome of the measurement on the type $X$ subsystem corresponds to an effect $_X(e|$, and a particular outcome of the measurement on the $Y$ subsystem corresponds to an effect $_Y(e|$, then the joint outcome corresponds to an effect $_{XY}(e| = _X(e| \otimes _Y(e|$.

Given a closed circuit, the joint probability for observing a particular collection of final pointer positions is given by contracting the various tensors to produce a real number. For example, consider an experiment corresponding to the closed circuit of Fig.~\ref{circuitfigure}. Reading from left to right: systems of types $A$ and $C$ are prepared; there is a transformation of the $A$ system into a system of type $B$; this is followed by a joint transformation of the $B$ and $C$ systems into systems of types $D$ and $E$; finally, a joint measurement is performed. The particular outcome of the experiment shown in Fig.~\ref{circuitfigure} corresponds to the pointers attaining positions $r_1, \ldots, r_5$. The theory assigns a probability to this outcome, given by:
$$
P(r_1\ldots r_5) = (G_{r_5}| \circ F_{r_4} \circ (T_{r_3}\otimes I ) \circ \left( |\sigma_{r_1})\otimes |\rho_{r_2}) \right),
$$
where $\circ$ denotes the action of a linear map on a vector, and $I$ is the identity operator on $V_C$.
\begin{figure}
$$
\begin{xy}
(65,11.5)*+{};(65,-10.5)*+{};
**\crv{(90,0)};
(13.8,12.5)*+{};(13.8,1.5)*+{};
**\crv{(-12,7)};
(13.8,0.5)*+{};(13.8,-10.5)*+{};
**\crv{(-12,-5)};
(24,7)*++[0][F-:<3pt>]{\text{$T_{r_3}$}}="1"; 
(47,1)*++++++[0][F-:<3pt>]{\text{$F_{r_4}$}}="2";
"1";(12.8,7)**\dir{-};
(8,7)*{\text{$\sigma_{r_1}$}};
(12.8,-5);(38,-5)**\dir{-};
(24,-8)*{\text{$C$}};
(12.8,12);(12.8,2)**\dir{-};
(8,-5)*{\text{$\rho_{r_2}$}};
(12.8,0);(12.8,-10)**\dir{-};
(15,4)*{\text{$A$}};
"1";(38,7)**\dir{-};
(56,7);(66,7)**\dir{-};
(61.3,4)*{\text{$D$}};
(61.3,-8)*{\text{$E$}}; 
(33,4)*{\text{$B$}};
(56,-5);(66,-5)**\dir{-};
(66,11);(66,-10)**\dir{-};
(71,0)*{\text{$G_{r_5}$}};
\end{xy}
$$
\caption{Example of a closed circuit in an operational theory. A particular outcome corresponds to the pointers on the devices attaining positions $r_1,\ldots, r_5$.}\label{circuitfigure}
\end{figure}

One can define a notion of \emph{causality} for theories in this framework: the probabilities of present experiments are independent of future measurement choices. It is shown in Ref.~\cite{Pavia1} that this requirement is equivalent to the existence of a unique deterministic effect for each type of system, denoted $_X(u|$, such that the following holds. First, for each measurement device, ${\sum_e {_X(e|} = {_X(u|}}$, where the sum is over the effects corresponding to outcomes of the device. Second, the norm of a state $|s)_X$ is given by $_X (u|s)_X$, and all states satisfy $_X (u|s)_X \leq 1$. Third, given a device with both input and output ports, the sum over the transformations corresponding to each pointer position must be a linear map that preserves the norm of the state. Note that consistent theories can be constructed with more than one deterministic effect, hence which violate causality \cite{acausal}. 

Finite dimensional quantum theory serves as an explicit example that illustrates the framework. Systems are associated with complex, finite dimensional Hilbert spaces, their type corresponding to the dimension of this space. States correspond to positive semi-definite operators acting on the underlying Hilbert space, with $V_X$ being the real vector space spanned by Hermitian operators. A measurement outcome is associated with a positive operator $E_X$ such that the corresponding effect is given by $\rho_X \rightarrow \mathrm{Tr} (E_X \rho_X)$. Quantum theory is of course causal: the unique deterministic outcome corresponds to the identity $I_X$, such that the positive operators for the different outcomes of the measurement must sum to $I_X$. The norm of a state $\rho_X$ is $\mathrm{Tr}(\rho_X)$. Transformations correspond to trace-non-increasing completely positive maps. A device with both input and output ports corresponds to a \emph{quantum instrument}, that is a set of trace-non-increasing completely positive (CP) maps (one for each pointer position) that sum to a trace-preserving CP map. It may be verified that tomographic locality holds in quantum theory. In particular, the state $\rho_{XY}$ of a composite system is a Hermitian operator acting on the tensor product of the underlying Hilbert spaces; the real vector space $V_{XY}$ spanned by such operators may be identified with $V_X \otimes V_Y$.

\color{black}

Theories different from quantum theory have also been studied in this framework. In the theory known as ``Boxworld'' \cite{PRbox, PR-trade-off}, for example, the simplest non-trivial type of system has a state defined via two binary-outcome measurements, $\{(x_a|\}$, where $x$ is a bit denoting the measurement setting and $a$ is a bit denoting the outcome. There are four possible pure states that can be prepared. Denoting these $|z,w)$, with $z,w\in\{0,1\}$, they satisfy $(0_b|z,w)=\delta_{bw}$ for measurement setting $0$, and $(1_b|z,w)=\delta_{bz}$ for measurement setting $1$. Multipartite states in the theory are defined such that aritrary non-signalling correlations can be produced, including, for example, the Popescu-Rohrlich correlations that maximally violate the CHSH inequality \cite{PRbox}. 
%That is, for a bipartite system $AB$, there exist states $|\rho_{PR})_{AB}$ such that
%$$(x_a|(y_b|\rho_{PR})_{AB}= \left\{ 
% \begin{array}{lr}
 %  \frac{1}{2}, \ \mathrm{if} \ a\oplus b= xy,\\
  %  0, \ \mathrm{otherwise}
 % \end{array}
%\right.
%$$
%where $\oplus$ represents addition modulo $2$. 
Boxworld satisfies both tomographic locality and causality \cite{PR-trade-off}.

Other interesting examples of non-quantum theories include the non-causal theory of Ref.~\cite{acausal}, and the theories investigated by Ref.~\cite{massar2015hyperdense}, in which the set of states of a single system corresponds to a Euclidean hyperball of dimension $n$. The toy theory of Ref.~\cite{Spekkens} may also be described by the operational framework.
%, with the $n=3$ case corresponding to the Bloch ball of quantum theory.  
\color{black} 
 
\subsection{Free and non-free theories}
 
In the usual definition of an operational theory \cite{Pavia1, Pavia2, Barrett-2007, Hardy-2011, LB-2014}, a theory specifies a set of laboratory devices, from which one can build closed circuits, and assigns a probability distribution over the outcomes of each closed circuit. Any closed circuit that can be built from the laboratory devices corresponds to a valid experiment. This means that there is a significant constraint on the structure of the theory, which is that all closed circuits must give rise to a valid probability distribution over outcomes. We refer to such theories as ``free'' operational theories \footnote{The idea behind the terminology is that an agent is free to build an experiment corresponding to any closed circuit they like, as long as devices are composed properly, i.e., types match.}.  

This work considers a more general definition of an operational theory, according to which a theory specifies a set of laboratory devices, and a set of \emph{allowed} closed circuits, which may be a proper subset of the set of all closed circuits that can be built using the laboratory devices. The interpretation is that it is only the allowed closed circuits that correspond to experimental procedures that can actually be performed. The theory must assign a valid probability distribution over the outcomes of any allowed closed circuit.  This definition is not unmotivated if one takes the viewpoint that a physical theory corresponds both to a consistent account of experimental data and to which experiments are implementable in principle. This is a significant generalization for the following reason. Given a closed circuit that is not in the allowed set, one may still contract the tensors associated with the device outcomes in order to produce a real number; but there is no constraint that this number has to be in the interval $[ 0,1 ]$.

Note that for tomographic locality to hold in a non-free theory, the set of allowed closed circuits must at least include a collection of experiments that are sufficient for local tomography to be carried out. In particular, for each state of a system of type $AB$, there should be allowed closed circuits involving local measurements on the subsystems such that, when the outcome probabilities for these circuits are known, the state is completely specified.

We also assume that the set of allowed closed circuits is itself closed under parallel composition, so that an experimenter may always choose to perform both of two valid experiments, independently of one another. In more detail, if $C_1$ is an allowed closed circuit, with outcomes $r_1,\ldots, r_k$, and $C_2$ is an allowed closed circuit, with outcomes $s_1,\ldots, s_l$, then the parallel composition is also an allowed closed circuit, corresponding to a valid experiment with outcomes $r_1,\ldots,r_k,s_1,\ldots,s_l$. We require that the outcome probabilities satisfy
\[
P_{C_1C_2}(r_1,\ldots,r_k,s_1,\ldots,s_l) = P_{C_1}(r_1,\ldots,r_k) P_{C_2}(s_1,\ldots,s_l),
\]
where $P_{C_1}$ denotes the outcome distribution that the theory assigns to the circuit $C_1$, similarly $P_{C_2}$ and the circuit $C_2$, and $P_{C_1C_2}$ denotes the outcome distribution for the parallel composition \footnote{Part of the reason for this assumption is that the idea of bounded-error computation makes little sense unless independent repetitions of a computation can be carried out, to verify the result, and reduce error probabilities close to zero.}.

\subsection{Computation}\label{sec:computationInGPT}

The class of ``yes/no'' problems that a quantum computer can solve efficiently is denoted by $\textbf{BQP}$ and much research has been concerned with how large this class is. At present, the tightest known upper bound is $\textbf{BQP}\subseteq\textbf{AWPP}$ \cite{AWPP}, where $\textbf{AWPP}$ is a classical complexity class, known to be contained in $\textbf{PP}$, hence in $\textbf{PSPACE}$ \cite{AWPP}. This class is formally defined in Methods.
%Sec.~\ref{awpp-def}. 

In order to define efficient computation in theories belonging to the framework introduced above, we need the notion of a (polynomially sized) uniform circuit family, and a condition for a circuit to accept an input. The following definition appeared in Ref.~\cite{LB-2014}. A polynomially sized uniform circuit family is a set of closed circuits $\{C_x\}$,
\color{black}
where $x$ ranges over finite-length bit strings and corresponds to the input to the problem,
\color{black}
such that:
\begin{enumerate} 
\item There is a gate set $\mathcal{G}$, consisting of laboratory devices, such that each circuit in the family is built from elements of $\mathcal{G}$.
\item The number of gates in the circuit $C_x$ is bounded by a polynomial in $|x|$.
\item For each type of system, there is a fixed choice of basis, relative to which transformations are associated with matrices. Given the matrix ${M}$ representing (a particular outcome of) a gate in $\mathcal{G}$, a Turing machine can output a matrix $\widetilde{{M}}$ with rational entries, such that $ | ({M} - \widetilde{{M}})_{ij} | \leq \epsilon$, in time polynomial in $\log(1/\epsilon)$. 
\item There is a Turing machine that, acting on input $x=x_1x_2\dots x_n$, outputs a classical description of $C_x$ in time bounded by a polynomial in $|x|$.
\end{enumerate}
\noindent
This produces\color{black}, for each $C_x$, \color{black} a description of an experiment, whose devices produce classical outcomes. Denoting the string of observed outcomes by $z$, the final output of the computation is given by an \emph{acceptor function} $a(z)\in \{0,1\}$, where there must exist a Turing machine that computes $a$ in time polynomial in the length of the input $|x|$. We say that a run of the experiment accepts an input string $x$ if the outcome string $z$ of the circuit $C_x$ satisfies $a(z)=0$. The probability that a computation accepts the input string $x$ is therefore given by
$$P_x({\mathrm{accept}}) \,=\! \sum_{z|a(z)=0}\!\!P(z),$$ where the sum ranges over all possible outcome strings $z$ of the circuit $C_x$ for which $a(z) = 0$.
 
\begin{definition} 
For an operational theory $\bold{G}$, let the class of problems that can be solved efficiently be denoted schematically $\bold{BGP}$. A language $\mathcal{L}$ is in the class $\bold{BGP}$ if the set of allowed circuits defined by $\bold{G}$ includes a poly-sized uniform circuit family, along with an efficient acceptor, such that
\begin{enumerate}
\item $x\in\mathcal{L}$ is accepted with probability at least $\frac{2}{3}$. 
\item $x\notin\mathcal{L}$ is accepted with probability at most $\frac{1}{3}$. \end{enumerate}  
\end{definition}

The constants in the above definition can be chosen arbitrarily as long as they are bounded away from a half by some inverse polynomial. The following theorem was proved for free theories in Ref.~\cite{LB-2014}, and follows without modification for non-free theories as well:
\begin{theorem}\label{previous}
For any operational theory $\bold{G}$ satisfying tomographic locality, $$\bold{BGP}\subseteq\bold{AWPP}.$$
\end{theorem} 
One might wonder if efficient quantum computation can achieve the bound of Theorem~\ref{previous}. In Appendix~\ref{appendix promise} we present a complexity-theoretic argument that may be considered evidence against such a possibility.

\subsection{Achieving the upper bound} \label{upperbound}

The main result of this work is the construction of an operational theory, satisfying causality and tomographic locality, that has \emph{exactly} the power of this upper bound. 
\begin{theorem}\label{mainresult}  
There exists an operational theory $\bold{G}$, satisfying causality and tomographic locality, such that $$\bold{BGP} = \bold{AWPP}.$$
\end{theorem}
Hence \textbf{AWPP}, despite having a slightly involved definition in terms of gap functions for non-deterministic Turing machines (see Methods),
%Sec.~\ref{awpp-def}
can be thought of much more intuitively as the class of problems efficiently solvable by tomographically local physical theories. 

An intuitive sketch of the proof of Theorem~\ref{mainresult} is as follows (for formal definitions and proofs, see Methods). First, we show that the class \textbf{AWPP} is perfectly captured by a quasi-probabilistic model of computation, defined via a Turing Machine with quasi-probabilistic transition weights with the constraint that the total weight of transitions from a given state must sum to $+1$. We refer to this model as an \emph{Affine Turing Machine}. See Fig.~\ref{Branch} for a schematic illustration. We then construct uniform poly-size circuits, in which the gates are certain affine transformations, which can simulate---and be simulated by---an Affine Turing Machine, and hence which also capture \textbf{AWPP}. 
\begin{figure}[t]
\begin{centering}
\scalebox{1}{
\begin{tikzpicture}
%[triangle/.style = { regular polygon, regular polygon sides=3 }, node rotated/.style = {rotate=180}]
 	\begin{pgfonlayer}{nodelayer}
		\node [style=black dot] (0) at (3, 5.75) {};
		\node [style=black dot] (1) at (1, 3.75) {};
		\node [style=new]  at (1.5, 5.5) {$2$};
		\node [style=new]  at (4.5, 5.5) {$-1$};
		\node [style=black dot] (2) at (5, 3.75) {};
		\node [style=new]  at (6.3, 3.5) {$5$};
		\node [style=new]  at (3.8, 3.45) {$-4$};
		\node [style=new]  at (7.1, 1.5) {$\frac{1}{2}$};
		\node [style=new]  at (9.2, 1.5) {$\frac{1}{2}$};
		\node [style=new]  at (10.2, -1) {\text{``Yes''}}; 
		\node [style=new]  at (8, -1) {\text{``No''}}; 
		\node [style=black dot] (3) at (-2, 1.75) {};
		\node [style=black dot] (4) at (1, 1.75) {};
		\node [style=black dot] (5) at (5, 1.75) {};
		\node [style=black dot] (6) at (8, 1.75) {};
		\node [style=black dot] (7) at (-4, -0.25) {};
		\node [style=black dot] (8) at (-2, -0.25) {};
		\node [style=black dot] (9) at (0, -0.25) {};
		\node [style=black dot] (10) at (2.5, -0.25) {};
		\node [style=black dot] (11) at (3.5, -0.25) {};
		\node [style=black dot] (12) at (6, -0.25) {};
		\node [style=black dot] (13) at (8, -0.25) {};
		\node [style=black dot] (14) at (10, -0.25) {};
		\node [style=new] (15) at (4.5, 5) {};
		\node [style=new] (16) at (1.5, 5) {};
		\node [style=new] (17) at (-1, 3) {};
		\node [style=new] (18) at (0.5, 2.5) {};
		%\node [style=new] (19) at (5.5, 2.5) {};
		%\node [style=new] (20) at (7.25, 3) {};
		%\node [style=new] (21) at (-3.5, 1.25) {};
		%\node [style=new] (22) at (-1.5, 0.75) {};
		%\node [style=new] (23) at (0.25, 1) {};
		%\node [style=new] (24) at (2, 1) {};
		%\node [style=new] (25) at (4, 1) {};
		%\node [style=new] (26) at (6, 1) {};
		%\node [style=new] (27) at (7.5, 1) {};
		%\node [style=new] (28) at (9.25, 1.25) {};
		%\node [style=point] (29) at (-8.25, -0.25) {};
		%\node [style=new] (30) at (-8.25, 1.75) {};
		%\node [style=small box] (31) at (-8.25, 3.75) {};
		%\node [style=copoint] (32) at (-8.25, 5.75) {};
		%\node [style=new] (33) at (-8.25, -1.25) {};
		%\node [style=new] (34) at (3, -1.25) {};
		%\node [style=new] (35) at (-7.25, 4.75) {}; 
		%\node [style=new] (36) at (3, 6.75) {};
		%\node [style=new] (37) at (-8.25, 6.75) {};
		%\node [style=point] (38) at (-6.5, -0.25) {};
		%\node [style=new] (39) at (-6.5, 1.75) {};
		%\node [style=small box] (40) at (-6.5, 3.75) {};
		%\node [style=copoint] (41) at (-6.5, 5.75) {};
		%\node [style=new] (42) at (-9.5, 0.75) {};
		%\node [style=new] (43) at (10.5, 0.75) {};
		%\node [style=new] (44) at (-9.5, 2.75) {};
		%\node [style=new] (45) at (10.5, 2.75) {};
		%\node [style=new] (46) at (-9.5, 4.75) {};
		%\node [style=new] (47) at (10.5, 4.75) {};
		%\node [style=new] (48) at (-7.5, 1.75) {};
		%\node [style=large box] (49) at (-7.4, 1.75) {};
	\end{pgfonlayer}
	\begin{pgfonlayer}{edgelayer} 
		\draw (0) to (1);
		\draw (1) to (3);
		\draw (3) to (7);
		\draw (3) to (8); 
		\draw (1) to (4);
		\draw (4) to (9);
		\draw (4) to (10);
		\draw (5) to (11);
		\draw (5) to (2);
		\draw (2) to (6);
		\draw (6) to (14);
		\draw (6) to (13);
		\draw (5) to (12);
		\draw (0) to (2);
		%\draw (32) to (31);
		%\draw (31) to (30);
		%\draw (30) to (29);
		%\draw (38) to (39);
		%\draw (39) to (40);
		%\draw (40) to (41);
		%\draw[dashed] (42) to (43);
		%\draw[dashed] (44) to (45);
		%\draw[dashed] (46) to (47);
	\end{pgfonlayer}
\end{tikzpicture} 
}
	\end{centering}
%\medskip
%\end{center}
	\caption{Schematic illustration of an Affine Turing Machine.} 
	\label{Branch}
\end{figure} 
This construction results in a collection of closed circuits, which correspond to the probability that the final result of the Affine Turing Machine is ``yes'' or ``no'' on inputs of different lengths. Finally, we construct a causal (non-free) operational theory, which contains the closed circuits necessary to simulate any Affine Turing Machine amongst its set of allowed circuits, along with sufficient additional circuits to ensure that tomographic locality holds. This proves Theorem~\ref{mainresult}.

\section*{Discussion}  \label{discussion} 

This work describes an operational theory, which satisfies causality and tomographic locality, such that the class of problems that can be efficiently solved by devices in that theory is exactly $\textbf{AWPP}$. This provides a converse to the results of Ref.~\cite{LB-2014}. To describe this construction, we introduce a new possibility: that of a ``non-free'' theory, in which the  possible transformations of systems are not necessarily closed under sequential and parallel composition.

An interesting feature of the $\textbf{AWPP}$-complete theory constructed in this paper is that it satisfies the principle of causality. The main result of \cite{LB-2014} was that for any theory satisfying tomographic locality, whether or not causality is satisfied, efficiently solvable computational problems are contained in \textbf{AWPP}. Taken together, these results show that computational circuits in any non-causal theory can always be efficiently simulated by circuits in a causal theory. Hence, in the landscape of general theories, ``acausality'' does not appear to be a resource for computation.

\begin{figure}
\centering 
\includegraphics[scale=0.25]{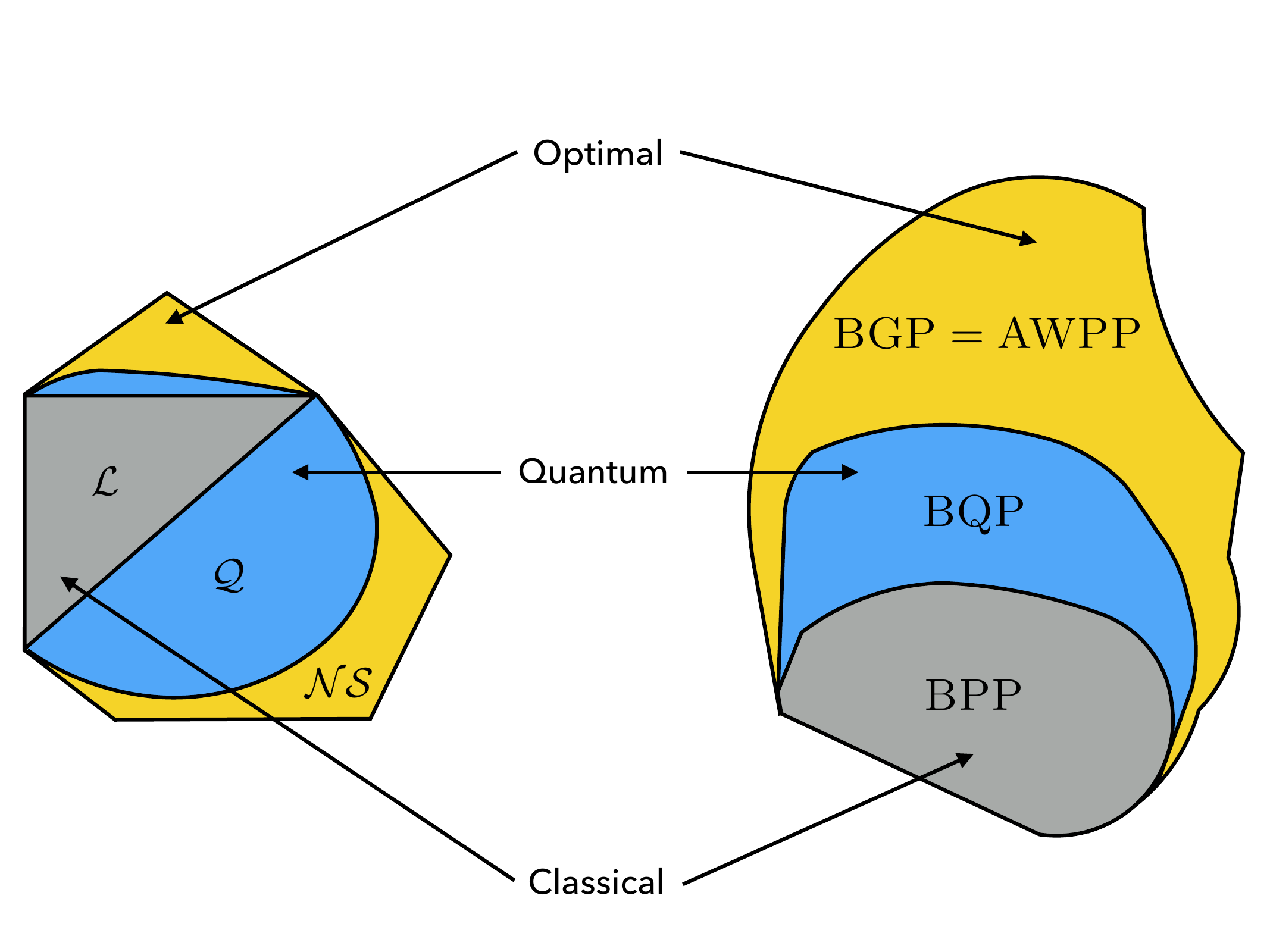}
\caption{Nonlocality versus computational power. The part of the figure on the left shows the set of all non-signalling correlations that could logically be obtained in a Bell test, with the sets of quantum and classical correlations strictly contained inside. The part of the figure on the right shows the computational complexity classes associated with theories that satisfy tomographic locality, with the theory presented in this paper saturating the whole of $\bold{AWPP}$, and the classes associated with the quantum and classical theories contained in this class. Note that in the case of computation, we can only conjecture that each of these containments is strict. For the containment of classical computation within quantum, see for example \cite{arkhipov, IQC} for evidence that it is strict.
}
\label{classes}
\end{figure}

Theorem~\ref{mainresult} is reminiscent of a result encountered when quantum correlations, obtained from measurements on entangled systems in a Bell-type experiment, are viewed in the context of the set of all non-signalling correlations \cite{Popescu-review}. Classical correlations are by definition local, and satisfy all Bell inequalities. Quantum correlations can be nonlocal, in the sense that they violate a Bell inequality, but the violation is limited by Tsirelson bounds \cite{Almost-quantum}. Operational theories can be constructed that produce stronger violations than is possible with quantum systems: for example, there exists a theory colloquially known as ``Boxworld'' \cite{Barrett-2007, PRbox} that allows all correlations consistent with the no-signalling principle. Similarly, when the computational power of tomographically local theories is considered, classical theories can be simulated by quantum theory, and it is believed that quantum computers can solve some problems efficiently that classical computers cannot. Here, we have shown the existence of an operational theory with the strongest possible computational power, and it is unlikely that quantum computers will be able to simulate this theory.
Fig. \ref{classes} schematically represents this analogy between the sets of correlations satisfying the no-signalling principle, and the computational complexity classes of theories satisfying tomographic locality, along with the quantum and classical cases for each.

Refs.~\cite{Quasi-prob-letter, Quasi-prob-phy, Samson} have, moreover, shown that methods employing quasi-probability distributions can simulate arbitrary non-signalling correlations. The quasi-probabilistic model of computation introduced here to build a theory with maximal computational power bears an intriguing resemblance to these approaches, providing another similarity between the set of all non-signalling correlations and the computational landscape of general theories.
 
Many attempts at providing reasonable physical principles that uniquely characterize the set of quantum correlations as a subset of the set of all non-signalling correlations have been made \cite{information-causality, Henson-2015, local-orthogonality, PR-van-Dam}. These principles, while not fully capturing the exact quantum boundary \cite{Almost-quantum}, have deepened our understanding of quantum correlations and provided connections between physical principles and information-theoretic advantages. Insights garnered from these connections led to the development of device-independent cryptography \cite{BHK-crypt, lee-hoban-2018, lee2018device}. So while investigating such connections has foundational interest, it has also been shown to have practical implications. 

It seems prudent to ask the analogous question for the set of tomographically local theories: can the class of efficient quantum computation be characterized by some set of physical principles? Such a characterization would deepen our understanding of quantum computation and may also be of practical relevance; if one uncovers the necessary and sufficient physical requirements for universal quantum computation one could design algorithms that optimally take advantage of them. The results presented in this paper provide one with the language and tools to pose these questions in a rigorous fashion

One approach to such a characterization would be to find the minimal set of physical principles that imply the quadratic speed-up over classical computation offered by Grovers search algorithm \cite{Nielsen-Chuang}. This speed-up is optimal for quantum computers \cite{Sub}, so any set of physical principles which imply it could be argued to capture some of the essence of quantum computation. Work in this direction has appeared in \cite{LS-2016,LS-2015,LS1-2015,NG}, where the quadratic lower bound to searching an unstructured database has been shown to hold for a large class of theories. 

Recently, methods have been proposed that make use of quasi-probability distributions to classically estimate the output of a quantum computer \cite{Quasi-prob-simulation}. These classical estimates converge on the true quantum output probabilities in a time quantified by the ``negativity'' of the quasi-probability distribution. The larger the negativity, the harder it is for a classical computer to estimate the output probability of a quantum computer. As we have provided an interpretation of the class \textbf{AWPP} in terms of quasi-probabilities, it would be interesting to determine if quantum algorithms can be constructed that estimate the output probability of this quasi-probabilistic computational model. In analogy with the classical estimation algorithms of \cite{Quasi-prob-simulation} the quantum algorithms may converge to the true output probability at a rate governed by the negativity of the quasi-probability distribution. Determining how hard it is for a quantum computer to simulate \textbf{AWPP} would provide a way to determining if quantum theory is powerful for computation in the landscape of general theories.   
     
Finally, the distinction introduced in this paper between free and non-free theories appears to be important for the study of computation in operational theories. Indeed, it is still an open question whether there exists a free theory whose computational power equals \textbf{AWPP}. The important distinction between free and non-free theories is that transformations in free theories are closed under composition, implying a bound on the set of states. This need not be the case in non-free theories. Could it be the case that a quantum computer can exploit this fact and efficiently simulate computation in all tomographically-local \emph{free} theories? If this conjecture holds, it could shed light on which physical features give rise to the quantum speed-up.

\section*{Methods}\label{main proof}

This section contains formal definitions, and the proof of Theorem~\ref{mainresult}.

\subsection{Definition of \textbf{AWPP}} \label{awpp-def}

Let $\Sigma$ be a finite set of symbols, \emph{e.g.}~$\Sigma = \{0,1\}$, and let $\Sigma^\ast$ be the set of all finite sequences over $\Sigma$ (commonly referred to as strings).
For a string $x \in \Sigma^\ast$, we let $|x|$ denote its length.
A \emph{gap function} over $\Sigma$ is a function $g: \Sigma^\ast \to \mathbb Z$ which computes the difference between the number of accepting branches and rejecting branches of some nondeterministic Turing machine $\mathbf N$, where $\mathbf N$ takes no more than $T(|x|)$ computational steps on input $|x|$ for some polynomial $T$ on whatever input $x$ it is given.

Fenner~\cite[Theorem~1.3]{Fenner-2003} characterized \textbf{AWPP} as the class of languages $L \subseteq \Sigma^\ast$ for which there is a gap-function $g: \Sigma^\ast \to \mathbb Z$ and a polynomial $p$, such that
\begin{subequations}
\label{eqn:FennerAWPPdefinition}
\begin{align}
  x \in L			\implies \tfrac{2}{3} \le \frac{g(x)}{2^{p(|x|)}} \le 1;
\\
	x \notin L	\implies 0 \le \frac{g(x)}{2^{p(|x|)}} \le \tfrac{1}{3}.
\end{align}
\end{subequations}
Combining this with \cite[Theorem~3.1]{Fenner-2003}, more generally we have $L \in \mathbf{AWPP}$ if and only if
\begin{subequations}
\label{eqn:AWPPdefinition}
\begin{align}
  x \in L			\implies \tfrac{2}{3} \le \frac{g(x)}{h(|x|)} \le 1;
\\
	x \notin L	\implies 0 \le \frac{g(x)}{h(|x|)} \le \tfrac{1}{3}.
\end{align}
\end{subequations}
for a gap-function $g$ and \emph{any} poly-time computable function $h: \mathbb N \to \mathbb N$.
While the original definition of \textbf{AWPP} \cite{FFKL} further required there to exist a gap-function $g$ and a poly-time computable function $h$ for any polynomial ${r: \mathbb N \to \mathbb N}$, satisfying either $g(x)\big/h(|x|) \in [0,2^{-r(|x|)}]$ or $g(x)\big/h(|x|) \in [1\,{-}\,2^{-r(|x|)},1]$, we instead use the characterizations of both Eqns.~\eqref{eqn:FennerAWPPdefinition} and~\eqref{eqn:AWPPdefinition} in our results.

\subsection{Affine Turing Machines}  
\label{AffTM}

We define an \emph{Affine Turing Machine} (AffTM) to be a non-deterministic Turing Machine, in which every transition has an associated  real-valued (possibly negative) \emph{weight}. \color{black} The weights for a given machine are constant throughout the computation, and should be thought of as defined by the transition function. \color{black} The weight of a given computational branch is then the product of the weights of the transitions involved.	
We require that for each symbol being read, the total weight of transitions from a given (non-halting) state is $+1$.
In this article we consider only rational transition weights, but expect that similar results would obtain for algebraic real coefficients.

We interpret AffTMs as a model of quasi-probabilistic computation, as follows.
Given an AffTM $\mathbf M$ whose branches all halt in in a finite number of steps, the \emph{acceptance weight} $\alpha_{\mathbf M}(x)$ of $\mathbf M$ on an input $x$ is the total weight of the accepting paths on input $x$.
An AffTM $\mathbf M$ is \emph{proper} if $0 \le \alpha_{\mathbf M}(x) \le 1$ for all inputs, and that it \emph{decides a language $L$ with bounded error} if furthermore $\tfrac{2}{3} \le \alpha_{\mathbf M}(x) \le 1$ for $x \in L$, and $0 \le \alpha_{\mathbf M}(x) \le \tfrac{1}{3}$ for $x \notin L$.

An AffTM is \emph{efficient} if the number of computational steps it takes in any computational path on any input $x$ is bounded by some polynomial in $|x|$. The first step towards Theorem~\ref{mainresult} is to establish the following:
\begin{lemma}\label{afftmlemma}
The class of languages decided with bounded error by some efficient AffTM is equal to $\bold{AWPP}$.
\end{lemma}
The proof of this result is contained in following two sections. 

\subsubsection{Solving \textbf{\upshape AWPP} problems with an affine Turing machine}  

For $L \subseteq \textbf{AWPP}$, let $g: \Sigma^\ast \to \mathbb Z$ be a gap-function satisfying Eqns.~\eqref{eqn:FennerAWPPdefinition} for some polynomial $p$.
Also let $\mathbf N$ be the non-deterministic Turing machine whose accepting/rejecting branches determine the gap-function $g$, and $T$ be the polynomial bounding the number of computational steps of $\mathbf N$ on its input.
By standard results~\cite{FFKL}, we may require that $\mathbf N$ have the same number of non-deterministic transitions at each step, which we denote by $N \ge 1$, and that all computational branches of $\mathbf N$ have the same length on input $x$.
We suppose that each transition of $\mathbf N$ is associated with some label $\ell \in \{1,2,\ldots,N\}$: the computational branches of $\mathbf N$ are then in one-to-one correspondence with sequences $\{1,2,\ldots,N\}^{T(|x|)}$.
We may then consider an AffTM $\mathbf M$ which simulates $\mathbf N$, in the following sense:
\begin{enumerate}
\item
	$\mathbf M$ first makes $T(|x|)$ non-deterministic transitions, writing a sequence of symbols $\beta_1, \beta_2, \ldots, \beta_{T(|x|)} \in \{0,1,2,\ldots,N\}$ on the tape to produce a string $\beta \in \{0,1,2,\ldots,N\}^{T(|x|)}$.
	The weights of these transitions are $+1$ for each choice $\beta_t \ne 0$, and $(1-N)$ for each choice $\beta_t = 0$, so that the transition weights sum to $+1$.
\item
	In branches with one or more symbols $\beta_t = 0$, $\mathbf M$ transitions deterministically with weight $+1$ to a state \mbox{\textsc{reject}}.
	All other branches of $\mathbf M$ have weight $+1$ and record a string $\beta \in \{1,2,\ldots,N\}^{T(|x|)}$ indexing some computational branch of $\mathbf N$.
	In these branches, $\mathbf M$ simulates the computational branch of $\mathbf N$ whose transitions are indexed by $\beta$.
\item
	For any branch in which the simulation of $\mathbf N$ rejects, $\mathbf M$ makes a non-deterministic transition to a state \mbox{\textsc{dampen}} with weight $-1$, and to the \mbox{\textsc{reject}} state with weight $+2$.
	For the branches in which the simulation of $\mathbf N$ accepts, $\mathbf M$ transitions deterministically to \mbox{\textsc{dampen}} with weight $+1$.
\item
	From the state \mbox{\textsc{dampen}}, $\mathbf M$ makes a sequence of $p(|x|)$ non-deterministic transitions with weight $\tfrac{1}{2}$, in which it writes bits $\delta_1, \delta_2, \ldots, \delta_{p(|x|)}$ on the tape, thereby sampling a string $\delta \in \{0,1\}^{p(|x|)}$ uniformly at random.
	If $\delta = 11\cdots1$, $\mathbf M$ transitions to an \mbox{\textsc{accept}} state; in all other branches it transitions to the \mbox{\textsc{reject}} state.
\end{enumerate}
By the construction of the branch weights, $\mathbf M$ is an AffTM; and as the number of transitions that $\mathbf M$ makes is $O(T + p)$, it is efficient.
By construction, the total weight of the branches which transition to the \mbox{\textsc{dampen}} state is $g(x)$; sampling the string $\delta \in \{0,1\}^{p(|x|)}$ and rejecting unless $\delta = 11\cdots1$ ensures that the acceptance weight is $\alpha_{\mathbf M}(x) = g(x)/2^{p(|x|)}$.
By hypothesis, this is bounded between $0$ and $1$, is at least $\tfrac{2}{3}$ if $x \in L$, and is at most $\tfrac{1}{3}$ otherwise.
Thus $\mathbf M$ decides $L$ with bounded error.

\subsubsection{Simulating an Affine Turing Machine in \textbf{\upshape AWPP}} 

Suppose that $\mathbf M$ is a proper and efficient AffTM which has transitions with rational weights. \color{black} Let $M$ be the common denominator of the (finite set of) transition weights of $\mathbf{M}$, and let $T \in O(\mathop{\mathrm{poly}}(n))$ be the running time of $\mathbf M$ on an input of length $n$. Let $m > 0$ be an integer, chosen such that $m \in O(T)$, and such that $2^m \ge M^T$ and $2^m \ge \bigl(|u| M)^T$ for all transition weights $u$ of $\mathbf M$. \color{black}We may obtain an \textbf{AWPP} algorithm to approximately simulate $\mathbf M$, as follows. We define a non-deterministic machine $\mathbf N$, which simulates $\mathbf M$ in the following sense.
\begin{enumerate}
\item
	The machine $\mathbf N$ reserves some space on the tape to represent some weight $\Omega \in \mathbb{Q}$ for each branch.
	We call this the \emph{recorded weight} of the branch.
	
\item
	Consider a transition made by $\mathbf M$, with weight $u = U/M$.
	To simulate this transition, the machine $\mathbf N$ replaces the recorded weight $\Omega$ with $\Omega' := U \Omega$, and \color{black}then \color{black}simulates the actions (writing of symbols and movement of the tape head) performed by $\mathbf M$ in the original transition.
	
\item
	Once $\mathbf N$ has simulated the final transition of $\mathbf M$, it non-deterministically samples a sequence of bits $a,b,c_0, c_1, \ldots, c_{m{-}1} \in \{0,1\}$.
	If $a = 1$, we negate $\Omega$ if and only if the simulated branch is one in which $\mathbf M$ rejects.

\item
	$\mathbf N$ determines whether to accept or reject, treating $c_{m{-}1} c_{m{-}2} \cdots c_1 c_0$ as the binary expansion of an integer $ 0 \le C < 2^m$, as follows.
	\begin{itemize}
	\item
		If $C \ge |\Omega|$, we reject if $b = 0$, and accept if $b = 1$.
	\item
		If $0 \le C < |\Omega|$, we reject if $\Omega < 0$, and accept if $\Omega > 0$.
	\end{itemize}
\end{enumerate}
Consider the gap function $g(x)$ of the machine $\mathbf N$.
From Step~4, it is clear that if $C \ge |\Omega|$ in any particular branch, $\mathbf N$ accepts and rejects with equal measure, contributing nothing to $g(x)$.
The significance of the contribution of any simulated branch of $\mathbf M$ is then in proportion to its recorded weight in $\mathbf N$, which in absolute value is $2M^T$ times its weight in $\mathbf M$ (arising from the systemic failure to divide the recorded weight by $M$ at each of the $T$ transitions, and from the two values of $b$).
Let $\alpha_+(x)$ be the total weight of those accepting branches of $\mathbf M$ with positive weight, $\alpha_-(x)$ be the total (absolute value of) the weight of accepting branches with negative weight; and similarly for $\rho_+(x)$ and $\rho_-(x)$ for rejecting branches of positive and negative weight.
Then $\alpha(x) := {\alpha_+(x) - \alpha_-(x)}$ is the acceptance weight and $\rho(x) := {\rho_+(x) - \rho_-(x)}$ is the rejection weight of $\mathbf M$ on input $x$.
We decompose $g(x) = g_0(x) + g_1(x)$, where $g_0(x)$ is the contribution to the gap from branches in which $a = 0$, and $g_1(x)$ is the contribution to the gap from branches in which $a = 1$.
We then have
$$
  g_0(x) = 2M^T \Bigl[ \alpha_+(x) + \rho_+(x) - \alpha_-(x) - \rho_-(x) \Bigr] = 2M^T,
$$
as $\alpha(x) + \rho(x) = 1$.
In the branches where $a = 1$, the sign of the contribution from rejecting branches is negated, so that
$$
\begin{aligned}[b]
  g_1(x) &= 2M^T \Bigl[ \alpha_+(x) - \rho_+(x) - \alpha_-(x) + \rho_-(x) \Bigr]
  \\&= 2M^T \Bigl[ 2\alpha(x) - 1 \Bigr],
\end{aligned} 
$$
again using $\alpha(x) + \rho(x) = 1$. Then $g(x) = 4M^T \alpha(x)$, and for $h(n) = 4M^T$, we have $0 \le g(x)/h(|x|) \le 1$ as $\mathbf M$ is proper. 
Furthermore, if $\mathbf M$ decides a language $L$ with bounded error, then either ${\tfrac{2}{3} \le g(x)/h(|x|) \le 1}$ or ${0 \le g(x)/h(|x|) \le \tfrac{1}{3}}$ according to whether $x \in L$ or $x \notin L$; then $L \in \mathbf{AWPP}$ as well.

\subsection{Constructing affine circuits}  \label{affine circuits}

The next step towards Theorem~\ref{mainresult} is to construct a family of circuits that can simulate a proper, efficient AffTM. The construction of the circuits is based on that used by Yao in \cite{Yao} to construct quantum circuits that simulate a quantum Turing Machine (and also on that of \cite{BooleanCircuits1, BooleanCircuits2} for circuits that simulate a probabilistic Turing machine). As before, let \textbf{M} be a proper and efficient AffTM with alphabet $\Sigma$, set of states $Q$ and transition amplitudes $\delta(q,a,\tau,q',a')\in\mathbb{Q}$ with $\tau \in\{\leftarrow,\circ,\rightarrow\}$; the symbols $\leftarrow$, $\rightarrow$ and $\circ$ are interpreted as the tape head of the AffTM moving to the left, moving to the right, and remaining stationary. Here $\delta$ is the transition weight of \textbf{M} to change to state $q'$, print $a'$ on the tape and move according to $\tau$, if the machine is currently in state $q$ and reading $a$. The condition on the weights in order for \textbf{M} to be an AffTM is: $\sum_{\tau,q',a'}\delta(q,a,\tau,q',a')=1$ for all $q\in Q$, $a\in \Sigma$. 

We may denote any configuration of the AffTM by a real basis vector 
$$|s_{-t}, q_{-t}, a_{-t},\cdots, s_i, q_i, a_i, \cdots, s_{t}, q_{t}, a_{t}),$$ 
where the index $-t \le i \le t$ denotes the $i$th cell of the tape and $t$ is the run time of the AffTM (there are $2t+1$ cells, numbered from $-t$ to $t$). Here $s_i$ takes on value $0$ when the head is not at cell $i$, value $1$ when it is at cell $i$ and the transition step has not occurred and value $2$ when the head has just moved according to a transition and is now at cell $i$. Note that we can represent $s_i$ with two bits. The label $q_i$ denotes the internal state of the machine at cell $i$, so $q_i\in Q\cup\{\emptyset\}$, where $q_i=\emptyset$ if and only if $s_i=0$; and $a_i \in \Sigma$ denotes the alphabet character printed on cell $i$. It is clear that $\ell$ bits, where $\ell=2+\bigl\lceil\log(|Q|{+}1)\bigr\rceil+\bigl\lceil\log(|\Sigma|)\bigr\rceil$, are required to represent the information at each cell. One can thus think of these basis vectors as being encoded by strings in $\{0,1\}^{(2t+1)\ell}$.

The transitions made along any one branch are represented by a sequence of these vectors, where each element of the sequence is the configuration of the machine at a given moment in time. The full state of the AffTM corresponds to an affine combination of such configurations, and the evolution of the AffTM corresponds to affine transformations of these configurations in superposition.
We may then simulate the AffTM by a uniform family of affine circuits.

Here, an ``affine circuit'' (in analogy to quantum circuits) refers to an acyclic network of ``affine gates'', each of which represents an affine transformation acting on real vectors. We demand that the matrices corresponding to these affine transformations have entries (with respect to the standard basis) that can be computed efficiently, i.e.~in poly-time, by an ordinary Turing Machine. We also demand that the description of the circuit can be computed efficiently, and in particular that it contain only a polynomial number of gates.

\begin{figure}[t]
\centering
	\includegraphics[width=0.45\textwidth]{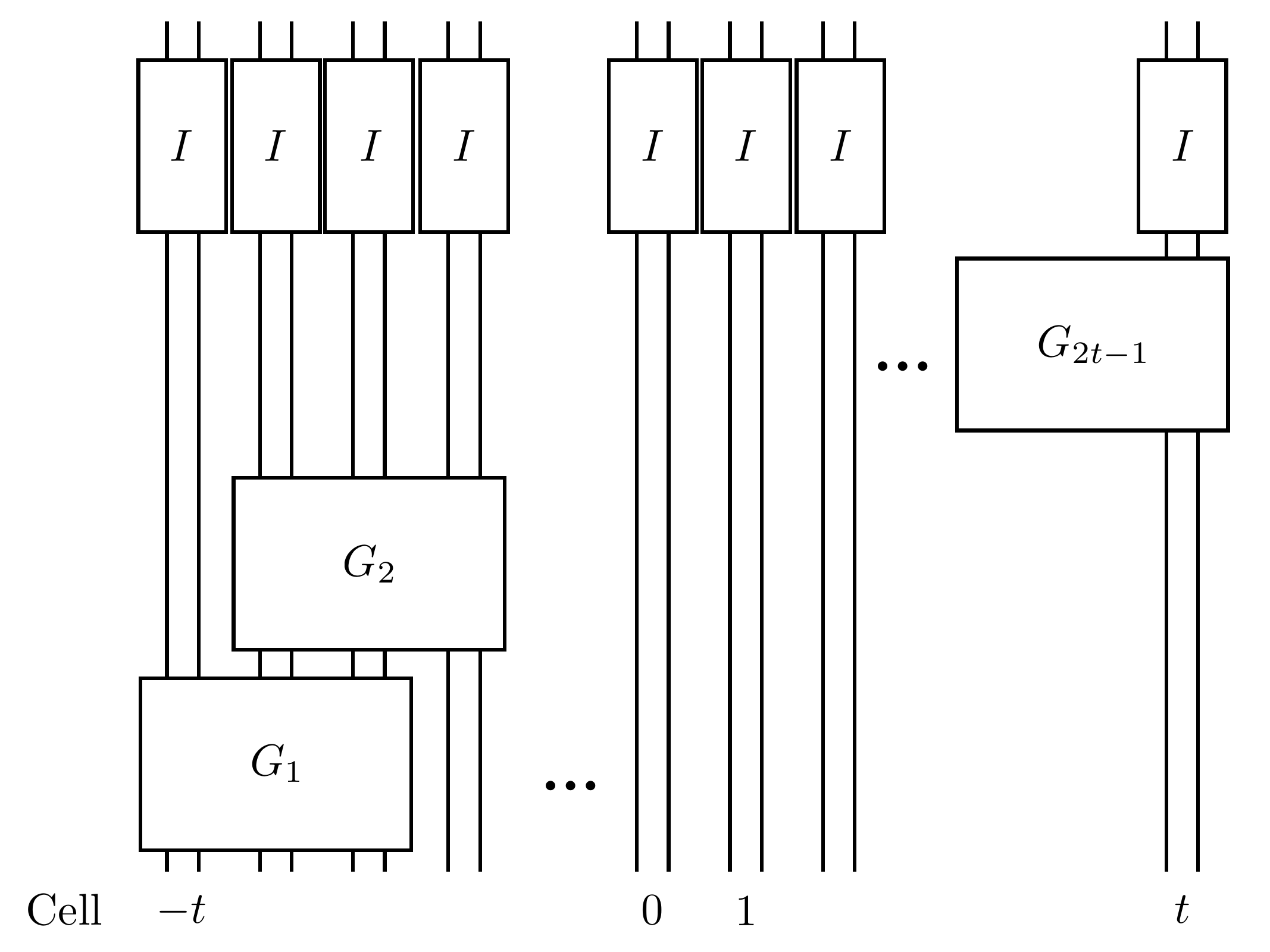}
	\caption{Sub-circuit $B$ simulating one transition of an affine Turing Machine $\mathbf M$.
	This circuit acts on $2t{+}1$ ``cells'', each consisting of $\ell$ wires and representing the contents of one cell of an affine Turing Machine $\mathbf{M}$, the location of the head, and the state of $\mathbf{M}$.} 
	\label{K}
\end{figure}

A specific affine circuit in this family will correspond to the concatenation of $t$ identical sub-circuits, which we denote by $B$. Each sub-circuit $B$ simulates one time-step of the AffTM $\mathbf M$. To construct these circuits, each tape cell of the AffTM is associated with $\ell = 2+\lceil\log(|Q|{+}1)\rceil+\lceil\log(|\Sigma|)\rceil$
wires in the circuit, which are sufficient to encode a tuple $(s_i, q_i, a_i) \in \{0,1,2\} \times \bigl(Q \cup \{\emptyset\}\bigr) \times \Sigma$, as described above. We build the sub-circuit $B$ with $3\ell$ input wires and $3\ell$ output wires, constructed from copies of two gates $G$ and $I$ as follows.
We first perform a cascading sequence of $2t-1$ copies of $G$ (whose behaviour we describe below), with each one shifted right by $\ell$ wires from the preceding one.
We then perform $2t+1$ copies of a gate $I$, in parallel, each acting on $\ell$ wires.
The gate $I$ acting on the $i$th cell changes the value of $s_i$ with value $2$ to $1$ and value $1$ to $2$, leaving a value of $s_i = 0$ alone.
It is clear that $I$ is an affine transformation and can be built using $O(t)$ gates whose function is to implement the change in $s_i$ for a specific $i$. We denote the $i$th instance of $G$ as $G_i$. See Fig.~\ref{K} for a pictorial representation of $B$. 

The intuitive idea behind this construction is as follows. The $3\ell$ inputs to $G$ should be thought of as describing the contents of three consecutive cells of the AffTM, including the information about the position of the head. We want $G$ to transform the contents of these cells if the head is at the middle cell and the transition step has not occurred (i.e.\ $s_i=1$ with $i$ being the middle cell) according to how the AffTM would transform the contents. Thus we design $G$ to act as follows: 
\begin{enumerate}
\item For all $v=|s_{i-1},q_{i-1},a_{i-1},s_i,q_i,a_i,s_{i+1},q_{i+1},a_{i+1})$ with $s_i\neq 1$, we have  
$G(v)=v,$
\item For $v'=|0,\emptyset, a_{i-1}, 1, q_i, a_i, 0, \emptyset, a_{i+1})$ we have  
$$\begin{aligned} 
G&(v') = \sum_{q',a'} \delta(q_i,a_i,\leftarrow,q',a')|2,q',a_{i-1}, 0,\emptyset,a', 0,\emptyset,a_{i+1}) \\
&+\sum_{q',a'} \delta(q_i,a_i,\circ,q',a')|0,\emptyset, a_{i-1}, 2,q',a',0,\emptyset,a_{i+1}) \\
&+\sum_{q',a'} \delta(q_i,a_i,\rightarrow,q',a')|0,\emptyset, a_{i-1},0,\emptyset,a',2,q',a_{i+1}).
\end{aligned}$$
\end{enumerate}   
\noindent
We can think of $G$ as a controlled affine transformation that does nothing if the input has $s_i\neq 1$ and performs the transition step of the AffTM otherwise.
(We may extend this to define $G|y) = |y)$ for any other basis state $|y)$, where $y \in \{0,1\}^{3\ell}$ does not encode a valid tuple $(s_{i{-}1},q_{i{-}1},a_{i{-}1},s_i,q_i,a_i,s_{i{+}1},q_{i{+}1},a_{i{+}1})$.)
As the configuration of the AffTM is an affine combination of vectors encoding tuples $\lvert s_{-t},q_{-t},a_{-t}, \ldots, s_{t},q_{t},a_{t})$, and as we have defined the action of $G$ (when tensored with the identity on cells on which it does not act) on all such vectors, extending linearly uniquely defines $G$'s action on all configurations of the AffTM. Note that some linear combination of vectors with $s_i\neq 1$ can lead to the same output as when $G$ is applied to a vector with $s_i=1$, so that $G$ may not be reversible. This may be expected, as affine transformations are not reversible in general; nor is there any requirement in the setting of operational theories to realise transformations reversibly.

We construct $B$ using a cascading sequence of $G$ gates, acting on the wires $1$ through $3\ell$ (representing cells $-t$, $-t+1$, and $-t+2$), then on the wires $\ell{+}1$ through $4\ell$, then $2\ell{+}1$ through $5\ell$, and so forth, as illustrated in Fig.~\ref{K}.
This in effect scans over the contents of the tape of the AffTM $\mathbf{M}$, doing nothing in most cases but simulating one of transition of $\mathbf M$ on the triple whose middle cell contains the head at the beginning of the transition.
The $I$ gates then flip the value of each $s_i$, so that the next simulation step can be performed. In this way, $B$ simulates one step of the AffTM.

We describe the initial state of the tape of $\mathbf M$ by setting $a_0a_1\cdots a_{n{-}1} = x_1 x_2 \cdots x_n$ (where $x \in \Sigma^\ast$ is the input of length $n$), and setting $a_i$ to the blank symbol for $i < 0$ and $i > n$.
We describe the initial head position of $\mathbf M$ by setting $s_0 = 1$ and $s_i = 0$ for $i \ne 0$; similarly we set $q_0 \in Q$ to the initial state of $\mathbf M$ and $q_i = \emptyset$ for $i \ne 0$.
This describes the initial state $|s_{-t},q_{-t},a_{-t},\ldots,s_t,q_t,a_t)$ which is the input to the affine circuit.
The run time of the simulated machine is $t$, so by concatenating $t$ instances of $B$ acting on the initial state, we obtain an affine circuit simulating the entire run of $\mathbf M$, producing a distribution $|\psi_x)$, which is an affine combination of basis vectors $|s'_{-t},q'_{-t},a'_{-t},\ldots,s'_t,q'_t,a'_t)$ representing the final configuration of all of the branches of the AffTM.

\begin{figure}[t]
\centering
	\includegraphics[width=0.5\textwidth]{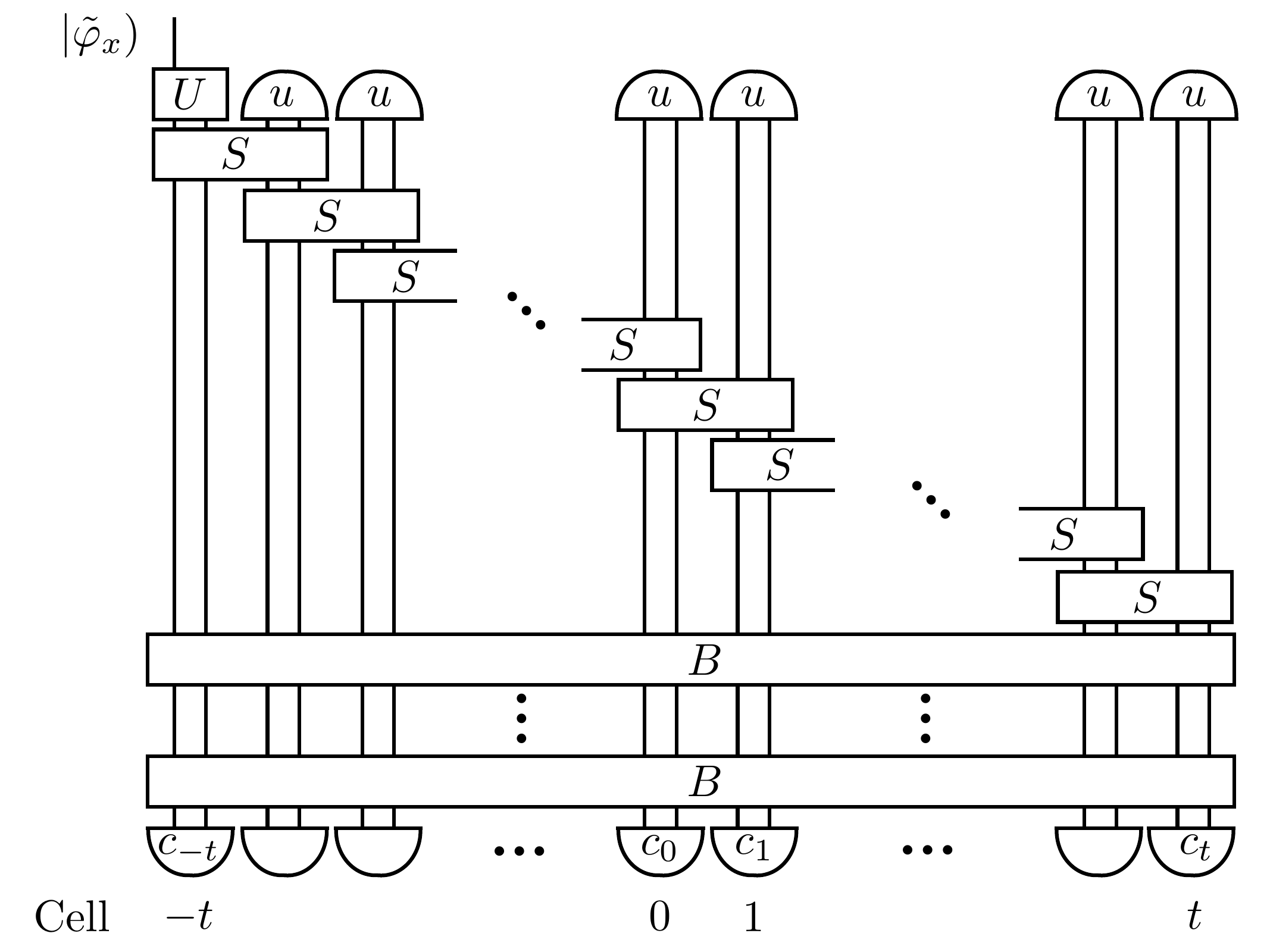}
	\caption{Schematic illustration of an affine circuit $M_n$ simulating an affine Turing Machine (AffTM) $\mathbf M$ on inputs of length $n$, which halts in time $t$.
	This includes $t$ copies of the gate $B$, each of which decomposes as the sub-circuit illustrated in Fig.~\ref{K} and simulates one transition of $\mathbf M$.
	The gate $S$ serves to simulate shifting the head of the AffTM to the leftmost of a pair of cells; the cascading sequence of $S$ gates serves to shift the head to the left-most cell in every branch of the computation, regardless of its final position when $\mathbf M$ halts.
	The preparations $|c_i)$ represent the initial configuration of the AffTM, preparing basis states $|c_i) = |s_i,q_i,a_i)$ where $s_0 = 1$ and $q_0$ is the AffTM initial state (and $s_i = 0$, $q_i = \emptyset$ for all $i \ne 0$), and where the symbols $a_i$ represent the symbols written on the $i$th cell of the tape.
	The final operations $(u| = \sum_{e} (e|$ on each cell serves to deterministically erase the information simulating the $i$th cell; the operation $U$ on the left-most cell similarly erases all information except for the distribution $|\tilde \varphi_x)$ representing the weights of the AffTM on the final internal states $|\mathrm{A})$ and $|\mathrm{R})$.}
	\label{AffineCircuit}
\end{figure}

As the position of the head in $\mathbf M$ in each branch may be different, we define another gate which will allow us to localise the final state of $\mathbf M$ in a definite subsystem.
We define a gate $S$ acting on $2\ell$ wires which transforms $|0,q'_i,a'_i,1,q'_{i{+}1},a'_{i{+}1}) \mapsto |1,q'_{i{+}1},a'_{i{+}1},0,q'_i,a'_i)$, and leaves all other basis states unchanged.
By performing a cascade of $S$ first on the wires $(2t{-}1)\ell{+}1$ through $(2t{+}1)\ell$ (representing cells $t{-}1$ and $t$), then on $2(t{-}2)\ell{+}1$ through $2t\ell$ (representing cells $t{-}2$ and $t{-}1$), and so forth, each standard basis state is mapped to one of the form $|1,\bar q,\bar a,s'_0,q'_0,a'_0,\ldots)$ for some $\bar q$ which is either the accept state $\mathrm A$ or reject state $\mathrm R$.
Acting on $|\psi_x)$, this cascade of $S$ gates produces a vector
$$|\varphi_x) = |1,\mathrm A)|\varphi_{\mathrm A,x}) + |1,\mathrm R)|\varphi_{\mathrm R,x}).$$
By the conditions on the acceptance weight of $\mathbf M$, the sum $w_{\mathrm A,x}$ of the coefficients of $|\varphi_{\mathrm A,x})$ satisfies either $w_{\mathrm A,x} \in [0,\tfrac{1}{3}]$ or $w_{\mathrm A,x} \in [\tfrac{2}{3},1]$; the same holds for the sum $w_{\mathrm R,x}$ of the coefficients of $|\varphi_{\mathrm R,x})$.
Applying the operator $_j(u| = {_j(0|} + {_j(1|}$ on all wires, except for the wires $3$ through $\lceil\log(|Q|+1)\rceil+2$ representing the final state $\mathrm A$ or $\mathrm R$ of $\mathbf M$, we then obtain a state 
\begin{equation}\label{affcircuitoutput}
|\tilde \varphi_x) = w_{\mathrm A,x}|\mathrm{A}) + w_{\mathrm R,x} |\mathrm{R})
\end{equation}
which is a distribution representing the probability with which $\mathbf M$ accepts $x$.
The entire affine circuit constructed in this way is illustrated in Fig.~\ref{AffineCircuit}.

The probability to accept is then just the factor in front of the basis state corresponding to the accepting configuration. We may thus simulate $\mathbf M$ by the $t$-fold application of $B$ on the initial configuration, followed by the cascade of $S$ gates and the application of unit effects described above.

\subsection{A tomographically local theory}\label{tomographThy}

The preceding section shows how to construct a collection of affine circuits that simulate a proper, efficient AffTM. In order to prove Theorem~\ref{mainresult}, this section constructs in turn a tomographically local operational theory, which can simulate a proper, efficient AffTM. It is important that tomographic locality is satisfied in order that Theorem~\ref{mainresult} serves as a converse to Theorem~\ref{previous}. As discussed in Ref.\cite{LB-2014}, theories that do not satisfy tomographic locality may have additional holistic degrees of freedom pertaining to composite systems. Without further constraint, there is nothing to stop such additional degrees of freedom enabling arbitrarily powerful computation. 

It is tempting to suppose that we need only construct an operational theory that includes the affine circuits of the last section. Each of the affine circuits outputs a state given by Eq.~(\ref{affcircuitoutput}), with accept and reject weights $w_{A,x}$ and $w_{R,x}$, and it follows from the premise that $\mathbf M$ is a \emph{proper} AffTM that $w_{A,x}, w_{R,x} \in [ 0,1 ] $. Hence if a circuit, representing an experiment in an operational theory, consists of the affine circuit, followed by a final measurement onto $|A)$ and $|R)$, the probabilities for the outcomes are at least guaranteed to be bounded by $0$ and $1$. Of course, closed circuits formed of arbitrary compositions of the same set of gates are not guaranteed to yield coefficients for measurement outcomes $\in [ 0,1 ] $, hence cannot be assumed to correspond to valid experiments. For this reason, the operational theory would be a non-free theory, with the set of allowed circuits containing those necessary for the simulation of proper, efficient AffTMs, but not allowing arbitrary rearrangements of gates. 

Even with the allowance of a non-free theory, however, it is not sufficient to define an operational theory as allowing exactly those circuits constructed above, along with a final accept/reject measurement. Without further structure, such a theory would simply be a theory of elaborate preparations of a $2$-dimensional system, whose states define probabilities for the acceptance and rejection outcomes. Additional structure is needed for the theory to satisfy tomographic locality, in such a way that states, transformations and effects correspond to the vectors and matrices that are involved in the construction of the affine circuits. 

The idea, therefore, is to allow circuits consisting of the initial part of one of the affine circuits, followed by measurements with outcomes corresponding to the basis states of each wire. If the effects were literally those dual to the basis states, this would suffice for tomographic locality; but the theory would not be well defined, because such effects would not in general yield sensible probabilities for outcomes. We therefore employ a trick, which is to allow only highly noisy versions of these measurements. If we additionally admix a small amount of noise with the final accept/reject measurement, then the theory can be shown to satisfy tomographic locality, to return sensible probabilities for measurement outcomes in all allowed circuits, and to be able to simulate a proper, efficient AffTM with bounded error. The precise construction is as follows.

Let $\{M_n\}_{n\ge 1}$ be the family of affine circuits, simulating a proper AffTM $\mathbf M$ on inputs of length $n \ge 1$. For each $n$, define types such that each wire gets a type $\nu_n$. This allows that the type of system involved can be distinct for each circuit in the family. From hereon, however, we consider a fixed $n$, suppressing the dependence of the type on $n$, and writing simply $\nu$. Define an \emph{initial segment} of $M_n$ to consist of any fragment that can be completed to the whole circuit $M_n$ by the post-composition of an appropriate sequence of gates (including, as a special case, $M_n$ itself). The closed circuits allowed by the theory consist, for each $n$, of an initial segment of $M_n$, followed by measurement devices attached to any dangling wires. 

First, for any system type $X$, there exists a measurement device realising the trivial measurement: the device pointer has only one position, which occurs with certainty. The outcome of this device corresponds to a deterministic effect, and the outcomes of any other measurement will correspond to effects that sum to the same deterministic effect, hence the theory is causal. For systems of type $\nu$, the deterministic effect is given by
$${_{\nu}(u|} = {_\nu(0|} + {_\nu(1|}.$$
For a composite system of type $X$, the deterministic effect ${_{X\!\!\:}(u|}$ is given by parallel composition. The deterministic effect may be appended to any dangling wire, following an initial segment of $M_n$.

Second, we define the measurements that enable local tomography. Define effects 
\begin{align*}
{_\nu(a_0|} &= p_\nu ({_\nu(0|}) + \frac{(1-p_\nu)}{2} ({_\nu (u|}), \\ 
{_\nu(a_1|} &= p_\nu ({_\nu(1|}) + \frac{(1-p_\nu)}{2} ({_\nu (u|}) , 
\end{align*}
where $p_\nu$ is a parameter to which we return below. These two effects satisfy
$${_\nu(a_0|}+ {_\nu(a_1|} \,= {_\nu(u|},$$
hence may correspond to the two outcomes of a binary measurement on a system corresponding to a single wire. This measurement may be appended to any dangling wire, following an initial segment of $M_n$.

Finally, there is the accept/reject measurement, which is a joint measurement defined on $\log(|Q|+1)$ systems of type $\nu$. Let the unit effect for such a collection of systems be $(u| = {_\nu{(u|^{\otimes \log(|Q|+1)}}}$, and let $(\mathrm{A}|$ and $(\mathrm{R}|$ denote the duals of the basis states $|\mathrm{A})$ and $|\mathrm{R})$, representing the accept and reject states (respectively) of the AffTM. The operational theory will allow a noisy version of the corresponding measurement, with effects given by:
\begin{align*}\label{accrejmeasure}
(e_{\mathrm{acc}}| &:= q (\mathrm{A}|  + (1-q) (\mathrm{R}|, \\
(e_{\mathrm{rej}}| &:=  q (\mathrm{R}|  + (1-q) (\mathrm{A}|, \\ 
(e_{\mathrm{none}}| &:= (u| - (e_{\mathrm{acc}}| -  (e_{\mathrm{rej}}|,
\end{align*}
with $q$ fixed independently of $n$, and essentially arbitrary as long as $1 > q > 1/2$. On pain of generating a disallowed circuit, this measurement cannot be appended to an arbitrary initial segment. The measurement can only be performed following an initial segment that is almost the whole of $M_n$, including at least all of the $S$ gates and the final $U$ gate (see Fig.~\ref{AffineCircuit}), and can only be performed on the $\log(|Q|+1)$ systems that are the output of the $U$ gate. 

The idea of this construction is that (separately for each value of $n$, the size of the problem input) the parameter $p_\nu$ can be chosen small enough that the measurements appearing in an allowed circuit always result in probabilities for outcomes that are bounded between $0$ and $1$. To see this, consider first those allowed circuits that include noisy tomographic measurements, but do not include the final accept/reject measurement. For these circuits, if $p_\nu =0$ then the outcomes of the noisy tomographic measurements each occur, independently, with probability $1/2$, regardless of the state. Now consider those allowed circuits that include the final accept/reject measurement, but where the final $_\nu (u|$ effect on one or more of the other wires has been replaced by noisy tomographic measurements. In this case, the probabilities for the accept and reject outcomes are bounded between $q$ and $1-q$, hence strictly between $1$ and $0$. It follows that if $p_\nu =0$, then the joint probability for either accept or reject, along with any sequence of outcomes for the tomographic measurements, is also strictly between $0$ and $1$. Now, in the theory under construction, there are only finitely many initial circuit fragments (in the partial construction of a single circuit on inputs of length $n$) on which to perform measurements. Continuity of the outcome probabilities in the effects therefore ensures that there exists a value $p_\nu >0$ such that joint outcome probabilities are contained in the interval $[0,1]$, for all circuits that can be constructed from systems of type $\nu$. Fixing such a value of $p_\nu$ results in noisy measurements that are sufficient for tomography on system $\nu$. 

Given a language decided by a poly-time, proper, bounded-error AffTM, the corresponding circuit family in the operational theory will accept yes-instances and reject no-instances with probabilities $\geq (1+q)/3$. If probabilities $\geq 2/3$ are required, they can be achieved by running several circuits in parallel. The final step in the proof of Theorem~\ref{mainresult} is to show how to combine the preceding constructions to describe an operational theory $\mathbf G$ not just for a single language in \textbf{AWPP}, but for the entire class.

As shown above, 
%in Sec.~\ref{AffTM}, 
every problem in \textbf{AWPP} can be solved with bounded error by a proper affine Turing machine (AffTM) which halts in polynomial time.
Conversely, any poly-time proper AffTM which has an acceptance weight either $\ge \tfrac{2}{3}$ or $\le \tfrac{1}{3}$ for all inputs, defines a language $L \in \mathbf{AWPP}$.
We then define a theory $\mathbf G$ which simply contains enough devices and system types to simulate every such AffTM, and only these AffTMs.
In this theory, each system type is parametrised by a (poly-time, proper, bounded-error) AffTM $\mathbf M$ and an input size $n \ge 1$; and each device is one of the sort described in the previous sections, also parameterised by $(\mathbf M,n)$.
The devices $G_{\mathbf M,n}$, $S_{\mathbf M,n}$, $I_{\mathbf M,n}$, and the various preparations and measurements for each system type, may then be used to construct circuits $C_{\mathbf M,n}$ to simulate the AffTM $\mathbf M$ on inputs of size $n$; and for each such $\mathbf M$, there will be a deterministic Turing machine $U$ which can generate $C_{\mathbf M,n}$ in $\mathrm{poly}(n)$ time.

To summarise: for any $L \in \mathbf{AWPP}$, there is a poly-time, proper AffTM $\mathbf M$ which decides $L$ with bounded error, which may be simulated by an affine circuit family $\{M_n\}_{n \ge 1}$.
This affine circuit family may be constructed uniformly, by the fact that it simulates an AffTM which halts in polynomial time.
The family $\{M_n\}_{n \ge 1}$ may itself be simulated by a uniform circuit family $\{C_{\mathbf M,n}\}_{n \ge 1}$ consisting of allowed experiments in the theory $\mathbf G$.
Then $\mathbf G$ is a non-free theory in which $\mathbf{AWPP} \subseteq \mathbf{BGP}$.
Together with Theorem~\ref{previous}, this concludes the proof of Theorem~\ref{mainresult}.

\subsection{On computation in non-free theories}\label{nonfreecomputation}

This section concludes by addressing a certain issue, which might arise with non-free theories: what if an agent can solve a hard problem (say, outside of \textbf{AWPP}) by simply observing whether a certain type of system exists in the universe or not? Or by simply observing whether a given circuit can be constructed or not? This would amount to a form of cheating, somewhat akin to the construction of non-uniform circuits in the classical or quantum cases. If such cheating were possible in a universe described by a non-free theory $\bold{G}$, this would not contradict the claim that $\bold{BGP} \subseteq \bold{AWPP}$, which is a formal mathematical theorem. But it would undermine the significance of the claim, since the definition of \textbf{BGP} could not be said to accurately capture the set of problems that an agent can efficiently solve by physical actions that the agent can do.

Concerning the first of these possibilities, our answer is that we have not said anything about how difficult it is to determine whether a given type of system exists in the universe or not: we can suppose, e.g., that the universe is infinite, and that given a classical description of an Affine Turing Machine, there is no step-by-step procedure that an agent can follow to determine if a corresponding type of system exists. Hence there is no easy way for an agent to solve the (uncomputable) problem of whether a given Affine Turing Machine is proper or not.

Concerning the second possibility, if a particular type of system is employed, the theory we construct does not allow a hard problem to be solved by finding out if a circuit is allowed or not. A closed circuit is allowed if it corresponds to an implementation of the corresponding Affine Turing Machine (or an initial segment thereof, with subsequent noisy measurements), and this is easy to check with a classical computation, hence the observation that a given circuit can or cannot be constructed cannot solve any harder problem. We argue therefore that we can rule out cheating in the theory described \footnote{More generally, one might require of a non-free theory something like the following: there exists a deterministic Turing machine, such that if the input is a description of a circuit, then on the promise that all the devices in the circuit exist in the universe, the machine decides in poly time whether the circuit is allowed or not. If the input is not a valid circuit, or contains devices that do not exist, then the output is unconstrained.}.

\paragraph*{Note added ---}While writing up the current work we became aware of the related but independent work \cite{CSW-2017}, on the characterization of \textbf{AWPP}. 

\section*{Data Availability}
Data sharing not applicable as no datasets were generated or analysed in the current manuscript.

\section*{Author Contributions}
All authors contributed equally to the current manuscript.

\section*{Competing interests}
The authors declare there are no competing interests.

\section*{Acknowledgements}
CML thanks J. Selby for useful discussions. We acknowledge support from the EPSRC National Quantum Technology Hub in Networked Quantum Information Technologies, an FQXi Large Grant and the Wiener-Anspach Foundation. This project and publication were made possible through the support of a grant from the John Templeton Foundation. The opinions expressed in this publication are those of the author(s) and do not necessarily reflect the views of the John Templeton Foundation.

\appendix

\section{Promise problems} \label{appendix promise}

One might wonder if efficient quantum computation can achieve the bound of Theorem~\ref{previous}. The following complexity-theoretic argument may be considered evidence against such a possibility.
\begin{theorem}\label{NP1} 
If $\bold{PromiseBQP}=\bold{PromiseAWPP}$, then 
$$\bold{NP}\subseteq\bold{BQP}\subseteq\bold{AWPP}.$$
\end{theorem}   
Here, the classes \textbf{PromiseBQP} and \textbf{PromiseAWPP} are \emph{promise} versions of the classes \textbf{BQP} and \textbf{AWPP}, meaning that they contain \emph{promise} rather than decision problems. 
A promise problem is a generalization of a decision problem, where the input is promised to belong to a subset of all possible inputs, so that there are disjoint subsets $\Pi_{\textrm{ACCEPT}},\, \Pi_{\textrm{REJECT}} \subseteq \Sigma^\ast$ of inputs to be accepted or rejected (respectively), but which do not exhaust the set of all inputs.
If an input belonging to neither $\Pi_{\textrm{ACCEPT}}$ nor $\Pi_{\textrm{REJECT}}$  is given to an algorithm for a certain promise problem, no requirements are placed on the output.

While, logically speaking, it could turn out that $\bold{BQP}=\bold{AWPP}$ without $\bold{PromiseBQP}=\bold{PromiseAWPP}$, 
this seems unlikely. Indeed, problems which are often regarded as complete for $\bold{BQP}$ or $\bold{AWPP}$ respectively, are in fact promise problems. Hence, $\bold{PromiseBQP}$ and $\bold{PromiseAWPP}$ can be loosely thought of as characterising the power of \textbf{BQP} and \textbf{AWPP} respectively. It is also believed unlikely \cite{Fenner-2003,GI, Sub} that \textbf{NP} is contained in either \textbf{BQP} or \textbf{AWPP}. Hence Theorem~\ref{NP1} can be regarded as evidence against the assertion that the computational power of quantum theory in the promise problem setting exactly equals \textbf{PromiseAWPP}, and this in turn may be regarded as evidence against the possibility that $\bold{BQP}=\bold{AWPP}$.

The proof of Theorem~\ref{NP1} is as follows.
\begin{proof}   
Recall that UNIQUE-SAT is the problem of deciding whether a given Boolean formula has exactly one satisfying truth assignment, or no satisfying assignment at all, promised that one of these is the case. It is known that UNIQUE-SAT is contained in \textbf{PromiseUP}, which is a subset of \textbf{PromiseAWPP} \cite{gap-def}. 

The Valiant-Vazirani theorem \cite{VV} says that if one has an efficient algorithm for solving UNIQUE-SAT in conjunction with the ability to perform random reductions, then one can solve any problem in \textbf{NP}. More precisely, the Valiant-Vazirani theorem says the standard Boolean Satisfiability Problem SAT can be randomly reduced to UNIQUE-SAT.

Now, if $\bold{PromiseBQP}=\bold{PromiseAWPP}$ then $\textrm{UNIQUE-SAT} \in \textbf{PromiseBQP}$, so that there is a uniform family of quantum circuits that solve an instance of the promise-problem UNIQUE-SAT (with no requirements made on inputs which do not respect the promise). However, a crucial point is that, as gates in quantum theory are closed under composition (in our terminology: quantum theory is a \emph{free} operational theory), the output of the algorithm will always result in sensible probabilities, regardless of the input. One can therefore perform the random reduction of Valiant-Vazirani in quantum theory (randomly generating an appropriate instance of SAT, and using this to generate an appropriate experiment of the sort that solves UNIQUE-SAT with bounded error), and run the algorithm many times on each input produced by the reduction to test whether it is a YES instance of UNIQUE-SAT. Performing this reduction many times enables the solution of SAT with bounded error in \textbf{BQP}.
It then follows that $\bold{NP}\subseteq \bold{BQP}$, which using Theorem~\ref{previous} gives $\bold{NP}\subseteq\bold{AWPP}$. 
\end{proof} 

One might wonder why the existence of a non-free theory satisfying $\bold{BGP}=\bold{AWPP}$ does not immediately imply $\bold{NP}\subseteq\bold{AWPP}$. The answer is that the theory we have constructed does not necessarily allow the efficient solution of $\bold{PromiseAWPP}$ problems, since the circuits required to simulate Affine Turing Machines that only have proper behaviour on a subset of inputs are not in the allowed set defined by the theory. 

One may then ask: why not construct an operational theory that \emph{does} contain circuits corresponding to simulations of the improper Affine Turing Machines that solve $\bold{PromiseAWPP}$ problems? In this case, the Valiant-Vazirani reduction does not go through, since the reduction assumes that it is possible to at least run the computation on inputs that do not satisfy the promise; attempting this in the operational theory must be disallowed since it may result in negative probabilities. On a related note, we would argue that such a theory should be excluded on the grounds discussed at the end of the Methods section.

\end{document}